\documentclass[prd,12pt]{article}
\usepackage{tikz}
\usepackage{amsmath,amssymb,graphicx,multirow,xspace,slashed,array,booktabs}
\usepackage[colorlinks=true,urlcolor=blue,anchorcolor=blue,citecolor=blue,filecolor=blue,linkcolor=blue,menucolor=blue,pagecolor=blue]{hyperref}
\usepackage[compress,numbers]{natbib}
\usepackage{subfigure, placeins}
\usepackage{booktabs}
\usepackage{blindtext, rotating}
\usepackage{afterpage}
\usepackage{enumitem}
\usepackage{ marvosym }
\usepackage{authblk} 
\usepackage{verbatim}
\usepackage{tikz-feynman}
\tikzfeynmanset{compat=1.0.0}
\usepackage{soul} %scratch out text
\usepackage[normalem]{ulem}
\usepackage{pifont}% http://ctan.org/pkg/pifont
\usepackage{booktabs}

\usepackage{floatrow}
\newfloatcommand{capbtabbox}{table}[][\FBwidth]

\usepackage[font=footnotesize,labelfont=bf]{caption}

\usepackage{lineno}

\allowdisplaybreaks

\long\def\symbolfootnote[#1]#2{\begingroup%
\def\thefootnote{\fnsymbol{footnote}}\footnote[#1]{#2}\endgroup}

\allowdisplaybreaks

\makeatletter
	
	\@addtoreset{equation}{section}
\makeatother

\newcommand{\newc}{\newcommand}
\newc{\gsim}{\lower.7ex\hbox{$\;\stackrel{\textstyle>}{\sim}\;$}}
\newc{\lsim}{\lower.7ex\hbox{$\;\stackrel{\textstyle<}{\sim}\;$}}
\newc{\gev}{\,{\rm GeV}}
\newc{\mev}{\,{\rm MeV}}
\newc{\ev}{\,{\rm eV}}
\newc{\kev}{\,{\rm keV}}
\newc{\tev}{\,{\rm TeV}}
\newc{\MHT}{$H_T^{\text{miss}}$}
\newc{\MET}{$\slashed{E}_T$}
\newc{\MTT}{$M_{T2}$}

\def\ln{\mathop{\rm ln}}

\newc{\mz}{M_Z}
\newc{\mpl}{M_*}
\newc{\mw}{m_{\rm weak}}
\newc{\nr}[1]{N^c_R{}_{#1}}

\def\beq{\begin{equation}}
\def\eeq{\end{equation}}
\newcommand{\bea}{\begin{eqnarray}\begin{aligned}}
\newcommand{\eea}{\end{aligned}\end{eqnarray}}
\def\bitem{\begin{itemize}}
\def\eitem{\end{itemize}}

\begin{document}
\baselineskip 0.6cm

\begin{titlepage}

\vspace*{-0.0cm}

\thispagestyle{empty}

\hfill OU-HET-1160

\begin{center}

\vskip 1cm

{\LARGE \bf A Large Muon EDM from Dark Matter}

\vskip 1cm

\vskip 1.0cm
{\large Kim Siang Khaw$^{1,2}$, Yuichiro Nakai$^{1,2}$, Ryosuke Sato$^3$, \\[1ex]
Yoshihiro Shigekami$^{1,2}$ and Zhihao Zhang$^{1,2}$}
\vskip 1.0cm
{\it
$^1$Tsung-Dao Lee Institute, Shanghai Jiao Tong University, \\ 520 Shengrong Road, Shanghai 201210, China \\
$^2$School of Physics and Astronomy, Shanghai Jiao Tong University, \\ 800 Dongchuan Road, Shanghai 200240, China \\
$^3$Department of Physics, Osaka University, Toyonaka, Osaka 560-0043, Japan}
\vskip 1.0cm

\end{center}

\vskip 1cm

\begin{abstract}
We explore a model of dark matter (DM) that can explain the reported discrepancy in the muon anomalous magnetic moment and predict a large electric dipole moment (EDM) of the muon. 
The model contains a DM fermion and new scalars whose exclusive interactions with the muon radiatively generate the observed muon mass. 
Constraints from DM direct and indirect detection experiments as well as collider searches are safely evaded. 
The model parameter space that gives the observed DM abundance and explains the muon $g-2$ anomaly leads to the muon EDM of {$d_{\mu} \simeq (4 \mathchar`- 5) \times 10^{-22} \, e \, {\rm cm}$} that can be probed by the projected PSI muEDM experiment. 
Another viable parameter space even achieves $d_{\mu} = \mathcal{O}(10^{-21}) \, e \, {\rm cm}$ reachable by the ongoing Fermilab Muon $g-2$ experiment and the future J-PARC Muon $g-2$/EDM experiment. 
\end{abstract}

\flushbottom

\end{titlepage}

%\tableofcontents

%#######################
\section{Introduction}
\label{intro}

The near-future discovery of the muon electric dipole moment (EDM) is highly expected by the reported discrepancy in the muon anomalous magnetic moment $(g-2)_\mu$, which may indicate the existence of physics beyond the Standard Model (SM) at or below the TeV scale \cite{Muong-2:2006rrc,Keshavarzi:2018mgv,Aoyama:2020ynm,Muong-2:2021ojo} (for a review, see ref.~\cite{Keshavarzi:2021eqa}), because the same new physics contribution naturally has the imaginary part which is relevant to the EDM. 
The current upper limit on the muon EDM is $|d_{\mu}| < 1.8 \times 10^{-19} \, e \, {\rm cm}$ (95\% C.L.)~\cite{Muong-2:2008ebm}. 
There is also a study on indirect bounds on the muon EDM by measuring EDMs of heavy atoms and molecules, which indicates $|d_{\mu}| < 2 \times 10^{-20} \, e \, {\rm cm}$~\cite{Ema:2021jds}. 
Moreover, the sensitivity to the muon EDM will be improved in the near future: the ongoing Fermilab Muon $g-2$ experiments~\cite{Chislett:2016jau} and projected J-PARC Muon $g-2$/EDM experiment~\cite{Abe:2019thb} will explore the muon EDM at the level of $10^{-21} \, e \, {\rm cm}$, while the Paul Scherrer Institute (PSI) muEDM experiment~\cite{Adelmann:2021udj,Sakurai:2022tbk,muonEDMinitiative:2022fmk} will reach the sensitivity of $6 \times 10^{-23} \, e \, {\rm cm}$. 

A fermion EDM $d_f$ is described by a dimension-five operator $\mathcal{L} \supset -\frac{i}{2} d_f \bar{f} \sigma_{\mu\nu} \gamma^5 f F^{\mu\nu}$ where $f$ is a Dirac fermion, $\sigma^{\mu\nu} \equiv \frac{i}{2} [\gamma^\mu, \gamma^\nu]$ and $F^{\mu\nu}$ is the photon field strength. 
Since this operator requires a chirality flip and left- and right-handed fermions carry different charges in the SM, we actually need a Higgs field insertion which makes the EDM operator effectively dimension-six. 
Therefore, a new physics contribution to a fermion EDM scales as $v_H/M^2$ where $v_H$ and $M$ denote the Higgs vacuum expectation value (VEV) and a new physics mass scale, respectively. 

To estimate the expected size of the muon EDM, we can consider four classes of new physics that generate the muon EDM as well as the anomalous magnetic moment~\footnote{The similar classification has been presented for the case of the electron EDM in ref.~\cite{Cesarotti:2018huy} (see also ref.~\cite{Nakai:2016atk}).}:
\begin{itemize}
\item {\bf Spurion approach.}
The chirality flip required to generate the EDM operator is provided by the muon Yukawa coupling $y_\mu$ or some coupling proportional to $y_\mu$. 
When the muon EDM $d_\mu$ is generated at the $k$-loop level, we expect 
\begin{align}
d_\mu \sim \delta_{\rm CPV} \left( \frac{\lambda^2}{16\pi^2} \right)^k \frac{m_\mu}{M^2} \, .
\end{align}
Here, $\delta_{\rm CPV}$ and $\lambda$ represent the size of CP-violating phases and couplings involved in the loop, and $m_\mu$ is the muon mass. 
Models in this class have been discussed in refs.~\cite{Ibrahim:1997gj,Feng:2003mg,Feng:2006ei,Feng:2008cn,Zhao:2014vga,Su:2022vju,Nakai:2022vgp}. 

\item {\bf Flavor changing approach.}
If the muon is converted to the tau lepton by a lepton flavor violating (LFV) interaction, the chirality flip can be provided by the tau Yukawa coupling $y_\tau$. 
In this case, we find
\begin{align}
d_\mu \sim \delta_{\rm CPV} \frac{y_{\mu\tau}^2}{\lambda^2} \left( \frac{\lambda^2}{16\pi^2} \right)^k \frac{m_\tau}{M^2} \, ,
\end{align}
with a LFV coupling $y_{\mu\tau}$ and the tau lepton mass $m_\tau$. 
Refs.~\cite{Hiller:2010ib,Omura:2015xcg,Abe:2019bkf,Hou:2021zqq} have explored models in this class. 
Note that if the model has a scalar leptoquark with appropriate charge assignment, the chirality flip can be picked up from a quark Yukawa coupling, e.g., the top Yukawa coupling. 
Moreover, there is an enhancement due to the color factor $N_C = 3$. 
Refs.~\cite{Cheung:2001ip,Arnold:2013cva,Dorsner:2016wpm,Dekens:2018bci,Altmannshofer:2020ywf,Babu:2020hun,Crivellin:2021rbq} have explored such a possibility in the context of the electron EDM and ref.~\cite{Crivellin:2018qmi} presented a general discussion for the case of the muon $g-2$ and EDM. 
In addition, a model with extra vector-like leptons also has a possibility to predict a large muon EDM~\cite{Hiller:2020fbu,Hamaguchi:2022byw} due to the chirality flip on a heavy lepton line. 

\item {\bf Radiative stability approach.}
New physics that produces the muon EDM also generates the muon mass by removing the attached photon. 
When we just assume that such a contribution to the muon mass does not exceed the correct value, the size of the muon EDM is expected to be
\begin{align}
d_\mu \sim \delta_{\rm CPV}  \frac{m_\mu}{M^2} \, ,
\label{eq:roughEDMrad}
\end{align}
because the same loop factor and coupling $\lambda$ are shared by the generated muon mass and EDM. 

\item{\bf Tuning approach.}
If the muon mass generated by new physics that produces the muon EDM exceeds the correct value, a fine-tuning is required. 
This (unlikely but logical) possibility allows us to obtain a very large muon EDM which is bounded by
\begin{align}
d_\mu \sim \delta_{\rm CPV} \lambda \left( \frac{\lambda^2}{16\pi^2} \right)^k \frac{v_H}{M^2}
\lesssim \delta_{\rm CPV} \frac{4 \pi v_H}{M^2} \, ,
\end{align}
for $\lambda \lesssim 4\pi$. 
\end{itemize}
Table~\ref{tab:muon EDM prospects} shows mass scales of new physics that produce the muon EDM at the one/two-loop level probed by the projected PSI muEDM experiment~\cite{Adelmann:2021udj,Sakurai:2022tbk,muonEDMinitiative:2022fmk} (the ongoing Fermilab Muon $g-2$ experiment~\cite{Chislett:2016jau} and the future J-PARC Muon $g-2$/EDM experiment~\cite{Abe:2019thb}). 
Aside from the tuning approach, the table indicates that the radiative stability approach generates the largest muon EDM and its near-future measurements can probe mass scales larger than the TeV scale. 
The present paper explores this fascinating possibility for the first time through the study of a concrete model to realize the radiative stability approach~\footnote{In ref.~\cite{Yin:2021yqy}, the author commented on the possibility of a large muon EDM in the context of the radiative stability approach.}. 

\begin{table}[!t]
\centering
\begin{tabular}{|c||c|c|}
\hline 
& 1-loop & 2-loop \\ 
\hline \hline
Spurion & 300 GeV (75 GeV)  &  16 GeV (4 GeV) \\ \hline
Flavor changing & 580 GeV (140 GeV) & 30 GeV (7 GeV) \\ \hline 
Radiative stability & \multicolumn{2}{c|}{5900 GeV (1400 GeV)} \\ \hline 
Tuning & \multicolumn{2}{c|}{$1.0 \times 10^6$ GeV ($2.5 \times 10^5$ GeV)} \\ \hline 
\end{tabular}
\vspace{0.3cm}
\caption{Mass scales of new physics that produce the muon EDM at the one/two-loop level probed by the projected PSI muEDM experiment~\cite{Adelmann:2021udj,Sakurai:2022tbk,muonEDMinitiative:2022fmk} (the Fermilab Muon $g-2$ and J-PARC Muon $g-2$/EDM experiments~\cite{Chislett:2016jau,Abe:2019thb}) for each approach presented in the main text. 
Here, we assume $\lambda \approx 0.65$ of around the weak coupling constant and $y_{\mu\tau} \approx 0.3$ which is roughly the maximum value allowed by the measurement of the branching ratio of $h \to \mu \tau$ (see, e.g., ref.~\cite{Omura:2015xcg}). 
For the leptoquark model, the mass scale is enhanced by $(y_{\mu t} / y_{\mu \tau})\sqrt{N_C m_t / m_{\tau}} \approx 57 y_{\mu t}$ where $y_{\mu t}$ is the leptoquark coupling to the muon and the top quark.}
\label{tab:muon EDM prospects}
\end{table}

We consider a model of dark matter (DM) that can address the muon $g-2$ anomaly. 
A DM fermion and new scalars exclusively couple to the muon, which leads to the radiative generation of the muon mass. 
The model contains a new CP-violating phase and produces the muon EDM. 
We will find that the model parameter space to give the observed DM abundance and explain the muon $g-2$ anomaly leads to the muon EDM of {$d_{\mu} \simeq (4 \mathchar`- 5) \times 10^{-22} \, e \, {\rm cm}$} probed by the PSI muEDM experiment. 
Furthermore, it will be shown that another viable parameter space even achieves $d_{\mu} = \mathcal{O}(10^{-21}) \, e \, {\rm cm}$ reached by the Fermilab Muon $g-2$ and J-PARC Muon $g-2$/EDM experiments, which is consistent with the estimate of Table~\ref{tab:muon EDM prospects}. 

The rest of the paper is organized as follows. 
Section~\ref{models} starts with the description of our DM model and explores the mass spectrum. 
We then calculate the radiatively generated muon mass and coupling to the Higgs boson and the muon EDM as well as the anomalous magnetic moment. 
They are all induced at the one-loop level. 
We also discuss deviations of the muon couplings to the Higgs and $Z$ bosons from those of the SM. 
In section~\ref{DMpheno}, phenomenology of DM in our model is explored. 
Section~\ref{results} summarizes the independent parameters of the model, and then presents our results to identify the parameter space that gives the observed DM abundance and explains the muon $g-2$ anomaly and indicate the size of the muon EDM. 
In section~\ref{conclusion}, we give conclusions and discussions. 
Loop integrals and full one-loop expressions are summarized in appendices.

%#######################
\section{Model description}
\label{models}

Our DM model is based on models proposed in ref.~\cite{Baker:2021yli}, which radiatively generate the muon mass and explain the muon $g-2$ anomaly. 
The model contains a single Dirac fermion $\psi$ and two scalar fields $\phi, \eta$. 
Charge assignments for the relevant particles are shown in Table~\ref{tab:class1}. 
\begin{table}[!t]
\begin{center}
\begin{tabular}{|c||ccccccc|}
\hline
& $L_L^{\mu}$ & $\mu_R$ & $H$ & $\psi_L$ & $\psi_R$ & $\phi$ & $\eta$ \\ \hline \hline
$SU(2)_L$ & $\mathbf{2}$ & $\mathbf{1}$ & $\mathbf{2}$ & $\mathbf{1}$ & $\mathbf{1}$ & $\mathbf{2}$ & $\mathbf{1}$ \\
$Y$ & $- \frac{1}{2}$ & $-1$ & $\frac{1}{2}$ & $Y_{\psi}$ & $Y_{\psi}$ & $Y_{\psi} + \frac{1}{2}$ & $Y_{\psi} + 1$ \\ \hline
$L_{\mu}$ & $-$ & $-$ & $+$ & $+$ & $+$ & $-$ & $-$ \\
$X$ & $+$ & $+$ & $+$ & $-$ & $-$ & $-$ & $-$ \\ \hline
$S_a$ & $+$ & $-$ & $+$ & $+$ & $+$ & $+$ & $-$ \\ \hline
\end{tabular}
\caption{Charge assignments for the relevant particles. 
$L_L^{\mu}$ and $\mu_R$ represent the second generation of the left- and right-handed leptons and $H$ is the SM Higgs field. 
The hypercharge of $\psi_{L,R}$ is taken as $Y_{\psi} = 0$ in the present paper. 
$L_{\mu}$ and $X$ are $Z_2$ symmetries associated with the muon number and the exotic particle number, while $S_a$ is a softly broken $Z_2$ symmetry to forbid the tree-level muon Yukawa coupling.}
\label{tab:class1}
\end{center}
\end{table}
The present paper focuses on the case with $Y_{\psi} = 0$~\footnote{The model with $Y_{\psi} = -1$ has a singlet real scalar $\eta$, and a CP phase appears in the scalar sector. 
In this case, however, CP violating effects necessarily involve the SM Higgs VEV, and therefore, the muon EDM is suppressed when the exotic particle masses are set to be around TeV.}. 
We introduce two $Z_2$ symmetries $L_{\mu}, X$ associated with the muon number and the exotic particle number, respectively. 
The former symmetry~\footnote{The $L_{\mu}$ symmetry can be enhanced to a global $U(1)_{L_{\mu}}$ symmetry when $\lambda''_{H \phi}$ in Eq.~\eqref{eq:Vscl} is set to be zero. 
This value is irrelevant to our current analysis. 
Note that even if we do not have $U(1)_{L_\mu}$ symmetry, $B-3L_e$ number and $B-3L_\tau$ number are conserved (for a review, see, e.g., Ref.~\cite{Buchmuller:2005eh}) and there is no washout of baryon asymmetry in the early universe.} makes it possible to avoid severe constraints from lepton flavor violating processes, while the latter one stabilizes the lightest exotic particle which is identified as DM. 
In addition, we assume a softly broken $Z_2$ symmetry $S_a$ to forbid the tree-level muon Yukawa coupling. 
The charge assignments lead to the following terms in the Lagrangian:
\begin{align}
\mathcal{L} &\supset \left( - y_{\phi} \overline{L_L^{\mu}} \phi^{\dagger} \psi_R - y_{\eta} \overline{\psi}_L \eta \mu_R - m_D \overline{\psi}_L \psi_R - \frac{m_{LL}}{2} \overline{\psi}_L \psi^c_L - \frac{m_{RR}}{2} \overline{\psi^c_R} \psi_R + {\rm h.c.} \right) - V_{\rm scl} \, , \label{eq:Lagpart} \\
V_{\rm scl} &= \sum_{s = H, \phi, \eta} \left[ m_s^2 s^{\dagger} s + \frac{\lambda_s}{2} (s^{\dagger} s)^2 \right] + \lambda_{H \phi} (H^{\dagger} H) (\phi^{\dagger} \phi) + \lambda_{H \eta} (H^{\dagger} H) (\eta^{\dagger} \eta) + \lambda_{\phi \eta} (\phi^{\dagger} \phi) (\eta^{\dagger} \eta) \nonumber \\
&\hspace{1.2em} + \lambda'_{H \phi} (H^{\dagger} \phi) (\phi^{\dagger} H) + \left( a H \eta^{\dagger} \phi + \frac{\lambda''_{H \phi}}{2} (H^{\dagger} \phi)^2 + {\rm h.c.} \right) \, . \label{eq:Vscl}
\end{align}
Note that all couplings in $V_{\rm scl}$ and $y_{\phi, \eta}$ can be real and positive by field redefinitions, while one phase of $m_D$, $m_{LL}$, and $m_{RR}$ cannot be removed. 
In fact, a combination $m_{LL} m_{RR} / m_D^2$ is independent of phase rotations, and we define a physical phase in the model as
\begin{align}
\theta_{\rm phys} \equiv \frac{1}{2} \arg \left( \frac{m_{LL} m_{RR}}{m_D^2} \right) = \frac{1}{2} \Bigl( \theta_L + \theta_R - 2 \theta_D \Bigr) \, ,
\label{eq:CPphys}
\end{align}
where $\theta_{D, L, R}$ denote phases of $m_D$, $m_{LL}$, and $m_{RR}$, respectively. 
Since $\psi$ is singlet under the SM gauge symmetry, we can define the left- and right-handed Majorana fermions as $\psi_{L,R}^M \equiv \psi_{L,R} + (\psi_{L,R})^c$ with $\psi_{L,R} = P_{L,R} \psi$ and $(\psi_{L,R})^c \equiv i \gamma^2 (\psi_{L,R})^*$. 
We assume that the exotic scalars $\phi, \eta$ do not acquire nonzero VEVs. 
As a result, no mixing between $H$ and $\phi$/$\eta$ is induced, and hence we can parameterize the SM Higgs field $H$ as
\begin{align}
H = \begin{pmatrix}
G^+ \\
\frac{1}{\sqrt{2}} \left( v_H + h^0 + i G^0 \right)
\end{pmatrix} \, ,
\end{align}
where $v_H = 246.22$ GeV is the SM Higgs VEV, $G^+$ and $G^0$ are Nambu-Goldstone modes, and $h^0$ is the SM Higgs boson. 
Note that a minimization condition leads to 
\begin{align}
m_H^2 &= - \frac{1}{2} \lambda_H v_H^2 \, . \label{eq:mincond}
\end{align}
Below, we will present the mass spectrum of exotic particles and calculate the radiatively generated muon mass and coupling to the Higgs boson and the muon EDM as well as the anomalous magnetic moment. 
Deviations of the muon couplings to the Higgs and $Z$ bosons from those of the SM will be also discussed. 
Note that for the neutrino sector, we need a further extension to reproduce the correct neutrino mixing angles, due to the muon number symmetry. 
We discuss some possibilities of the extension in appendix~\ref{app:Neutrino}. 
We emphasize that such an extension does not affect our numerical results.

%#######################
\subsection{Mass spectrum of exotic particles}
\label{sec:exoticmasses}

From the Lagrangian \eqref{eq:Lagpart}, the mass matrix for $\psi_L$ and $\psi_R$ is
\begin{align}
- \frac{1}{2} \begin{pmatrix}
\overline{\psi}_L & \overline{\psi^c_R}
\end{pmatrix} \begin{pmatrix}
m_{LL} & m_D \\
m_D & m_{RR}
\end{pmatrix} \begin{pmatrix}
\psi_L^c \\
\psi_R
\end{pmatrix} + {\rm h.c.} = - \frac{1}{2} \begin{pmatrix}
\overline{\psi}_L & \overline{\psi^c_R}
\end{pmatrix} \mathcal{M}_{\psi} \begin{pmatrix}
\psi^c_L \\
\psi_R
\end{pmatrix} + {\rm h.c.} \, ,
\label{eq:MajoranaMassMatrix}
\end{align}
where $\mathcal{M}_{\psi}$ is a complex symmetric matrix and diagonalized by a unitary matrix $U_{\psi}$:
\begin{align}
\mathcal{M}_{\psi, {\rm diag}} = U_{\psi}^{\dagger} \mathcal{M}_{\psi} U_{\psi}^* \, , \quad U_{\psi} = \begin{pmatrix}
c_{\alpha} & s_{\alpha} e^{- i \tau} \\
- s_{\alpha} e^{i \tau} & c_{\alpha}
\end{pmatrix} \, .
\label{eq:U1}
\end{align}
Here, $c_\alpha \equiv \cos \alpha$ with mixing angle $\alpha$ and $\tau$ is real. 
In our analysis, we take $m_{LL}$ and $m_{RR}$ to be real and positive, while $m_D$ has a physical phase as $m_D = |m_D| e^{- i \theta_{\rm phys}}$. 
We then obtain mass-squared eigenvalues of $\mathcal{M}_{\psi}^{\dagger} \mathcal{M}_{\psi}$ as
\begin{align}
m_{\psi_1}^2 &= \frac{1}{2} \Bigl( m_{LL}^2 + m_{RR}^2 + 2 \left| m_D \right|^2 - \Delta m_{\psi}^2 \Bigr) \, , \\
m_{\psi_2}^2 &= \frac{1}{2} \Bigl( m_{LL}^2 + m_{RR}^2 + 2 \left| m_D \right|^2 + \Delta m_{\psi}^2 \Bigr) \, ,
\end{align}
where $\Delta m_{\psi}^2 \equiv m_{\psi_2}^2 - m_{\psi_1}^2$ is given by
\begin{align}
\Delta m_{\psi}^2 = \sqrt{\left( m_{LL}^2 - m_{RR}^2 \right)^2 + 4 \left| m_D \right|^2 \Bigl| m_{LL} e^{- i \theta_{\rm phys}} + m_{RR} e^{i \theta_{\rm phys}} \Bigr|^2} \, ,
\label{eq:defDelmpsi2}
\end{align}
with $\theta_{\rm phys}$ defined in Eq.~\eqref{eq:CPphys}. 
The mixing angle $\alpha$ and phase $\tau$ in $U_{\psi}$ are obtained as
\begin{align}
\sin 2 \alpha &= \frac{2 | m_D |}{\Delta m_{\psi}^2} \Bigl| m_{LL} e^{- i \theta_{\rm phys}} + m_{RR} e^{i \theta_{\rm phys}} \Bigr| \, , \\[0.5ex]
\tan \tau &= - \frac{m_{LL} - m_{RR}}{m_{LL} + m_{RR}} \tan \theta_{\rm phys} \, . \label{eq:tantau}
\end{align}
Due to the mass hierarchy, $m_{\psi_2}^2 > m_{\psi_1}^2$, we can focus on $0 \leq \alpha \leq \pi/2$. 
Note that physical predictions are unchanged for $\theta_{\rm phys} \to \theta_{\rm phys} + \pi$, and we focus on the range of $- \pi / 2 < \theta_{\rm phys} \leq \pi / 2$ in our analysis. 
$\psi_{L, R}$ can be described in terms of mass eigenstates $\psi_{1, 2}$ as
\begin{align}
\psi_L = \psi_1^c c_{\alpha} + \psi_2^c s_{\alpha} e^{- i \tau} \, , \quad \psi_R = - \psi_1 s_{\alpha} e^{- i \tau} + \psi_2 c_{\alpha} \, ,
\label{eq:MajoranamasseigenMF}
\end{align}
and the mass terms in Eq.~\eqref{eq:MajoranaMassMatrix} become
\begin{align}
- \frac{1}{2} \Bigl[ m_{\psi_1} \overline{\psi_1^M} \psi_1^M + m_{\psi_2} \overline{\psi_2^M} \psi_2^M \Bigr] \, ,
\end{align}
where $\psi_{1,2}^M \equiv \psi_{1,2} + \psi_{1,2}^c$ are Majorana fermions. 

In order to analyze the mass spectrum for exotic scalar fields, we parameterize them as
\begin{align}
\phi = \begin{pmatrix}
\phi^+ \\
\frac{1}{\sqrt{2}} \left( \sigma_{\phi} + i a_{\phi} \right)
\end{pmatrix} \, , \qquad \eta = \eta^+ \, .
\end{align}
From Eq.~\eqref{eq:Vscl}, the mass-squared matrices for charged and neutral scalars (in the basis of $(\phi^+, \eta^+)$ and $(\sigma_{\phi}, a_{\phi})$, respectively) are given by
\begin{align}
\mathcal{M}_{\pm}^2 &= \begin{pmatrix}
{\displaystyle M_{\phi}^2} & {\displaystyle \frac{a v_H}{\sqrt{2}}} \\[2.5ex]
{\displaystyle \frac{a v_H}{\sqrt{2}}} & {\displaystyle M_{\eta}^2}
\end{pmatrix} \, , \label{eq:M2pm} \\[1.5ex]
\mathcal{M}_0^2 &= \begin{pmatrix}
{\displaystyle M_{\phi}^2 + \frac{\lambda^+_{H \phi}}{2} v_H^2} & 0 \\[2.0ex]
0 & {\displaystyle M_{\phi}^2 + \frac{\lambda^-_{H \phi}}{2} v_H^2}
\end{pmatrix} \equiv \begin{pmatrix}
m_{\sigma_{\phi}}^2 & 0 \\
0 & m_{a_{\phi}}^2
\end{pmatrix} \, , \label{eq:M20}
\end{align}
where $M_{\phi, \eta}^2 \equiv m_{\phi, \eta}^2 + \frac{\lambda_{H \phi, H \eta}}{2} v_H^2$ and $\lambda^{\pm}_{H \phi} \equiv \lambda'_{H \phi} \pm \lambda''_{H \phi}$. 
Note that since all quartic couplings are positive, $m_{a_{\phi}}^2$ is always smaller than $m_{\sigma_{\phi}}^2$. 
Diagonalzation of $\mathcal{M}_{\pm}^2$ can be done by an orthogonal matrix as
\begin{align}
\mathcal{M}_{\pm, {\rm diag}}^2 = U_s^T \mathcal{M}_{\pm}^2 U_s, \quad U_s \equiv \begin{pmatrix}
c_{\theta} & s_{\theta} \\
- s_{\theta} & c_{\theta}
\end{pmatrix} \, .
\label{eq:Usdef}
\end{align}
The mass-squared eigenvalues and the mixing angle $\theta$ are 
\begin{align}
m_{\varphi_1^+}^2 &= \frac{1}{2} \left[ M_{\phi}^2 + M_{\eta}^2 - \sqrt{(M_{\phi}^2 - M_{\eta}^2)^2 + 2 a^2 v_H^2} \right] \, , \label{eq:eigenpm1} \\
m_{\varphi_2^+}^2 &= \frac{1}{2} \left[ M_{\phi}^2 + M_{\eta}^2 + \sqrt{(M_{\phi}^2 - M_{\eta}^2)^2 + 2 a^2 v_H^2} \right] \, , \label{eq:eigenpm2} \\
\sin 2 \theta &= \frac{\sqrt{2} a v_H}{m_{\varphi_2^+}^2 - m_{\varphi_1^+}^2} \, . \label{eq:mixanglepm}
\end{align}
Then, $\phi^{\pm}$ and $\eta^{\pm}$ can be described in terms of mass eigenstates $\varphi_{1,2}^{\pm}$ as
\begin{align}
\phi^{\pm} = \varphi_1^{\pm} c_{\theta} + \varphi_2^{\pm} s_{\theta} \, , \qquad \eta^{\pm} = - \varphi_1^{\pm} s_{\theta} + \varphi_2^{\pm} c_{\theta} \, .
\label{eq:scalarmasseigenMF}
\end{align}
Since the mass parameter $a$ can be set to be real and positive and $m_{\varphi_2^+}^2 - m_{\varphi_1^+}^2 > 0$, we can focus on $0 \leq \theta \leq \pi/2$.

%#######################
\subsection{Radiative mass and coupling of the muon}
\label{sec:muradandYmueff}

\begin{figure}[t]
\begin{center}
\begin{tikzpicture}
\begin{feynman}[large]
%%% leading contribution to effective Yukawa %%%
\vertex (a1) {\(\mu_R\)};
\vertex [right=1.5cm of a1] (b1);
\vertex [right=3cm of b1] (c1);
\vertex [right=1.1cm of c1] (d1) {\(\mu_L\)};
\vertex [above right=2.12132cm of b1] (f1);
\vertex [above right=1.5cm of f1] (g1) {\(h^0\)};
\vertex [below=2.2cm of f1] (l1) {\(\text{(a)}\)};
\diagram [medium] {
(a1) -- (b1) -- [edge label'=\(\psi_a\)] (c1) -- (d1),
(b1) -- [scalar, quarter left, edge label=\(\varphi_j^+\)] (f1) -- [scalar, quarter left, edge label=\(\varphi_i^-\)] (c1),
(f1) -- [scalar] (g1),
};
%%% leading contribution to dipole operators %%%
\vertex [right=1.5cm of d1] (a2) {\(\mu_R\)};
\vertex [right=1.5cm of a2] (b2);
\vertex [right=3cm of b2] (c2);
\vertex [right=1.1cm of c2] (d2) {\(\mu_L\)};
\vertex [above right=2.12132cm of b2] (f2);
\vertex [above right=1.5cm of f2] (g2) {\(\gamma\)};
\vertex [below=2.2cm of f2] (l2) {\(\text{(b)}\)};
\diagram [medium] {
(a2) -- (b2) -- [edge label'=\(\psi_a\)] (c2) -- (d2),
(b2) -- [scalar, quarter left, edge label=\(\varphi_i^+\)] (f2) -- [scalar, quarter left, edge label=\(\varphi_i^-\)] (c2),
(f2) -- [photon] (g2),
};
\end{feynman}
\end{tikzpicture}
\end{center}
\vspace{-0.5cm}
\caption{Feynman diagrams for (a) the effective Yukawa coupling of the muon and (b) dipole operators. 
The diagram for the radiative mass of the muon is obtained by eliminating the external photon line from the diagram (b).}
\label{fig:MFdiagrams}
\end{figure}
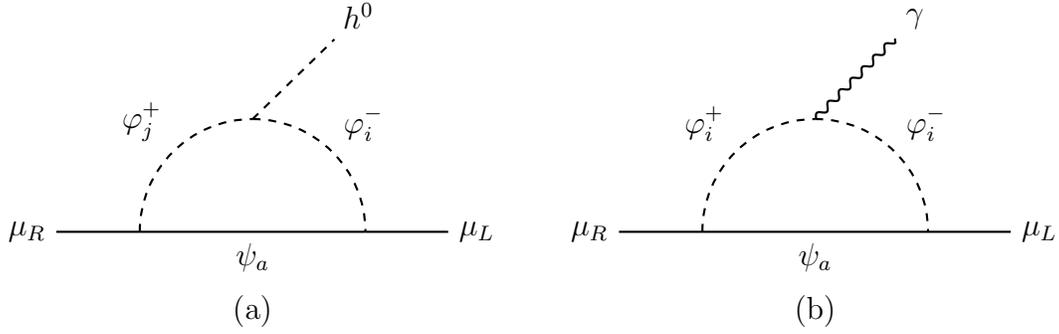

The mass and Yukawa coupling of the muon are induced by one-loop corrections. 
When we move to the mass basis for exotic particles according to Eqs.~\eqref{eq:MajoranamasseigenMF} and \eqref{eq:scalarmasseigenMF}, the relevant terms are written as 
\begin{align}
\mathcal{L} \supset \Bigl( - y_L^{i a} \bar{\mu}_L \varphi_i^- \psi_a - y_R^{i a} \overline{\psi_a^c} \varphi_i^+ \mu_R + {\rm h.c.} \Bigr) - \frac{A_{i j}}{\sqrt{2}} h^0 \varphi_i^- \varphi_j^+ \, ,
\label{eq:relevantterms}
\end{align}
where the explicit forms of $y_{L, R}^{i a}$ and $A_{i j}$ are summarized in Table~\ref{tab:yLRAij}. 
\begin{table}[th]
\begin{center}
\begin{tabular}{|c|cc||c|c|}
\hline
$(i, a)$ & $y_L^{i a}$ & $y_R^{i a}$ & $(i, j)$ & $A_{i j}$ \\ \hline
$(1, 1)$ & $- y_{\phi} c_{\theta} s_{\alpha} e^{- i \tau}$ & $- y_{\eta} s_{\theta} c_{\alpha}$                & $(1, 1)$ & $- a s_{2 \theta} + \sqrt{2} v_H \left( \lambda_{H \phi} c_{\theta}^2 + \lambda_{H \eta} s_{\theta}^2 \right)$ \\
$(1, 2)$ & $y_{\phi} c_{\theta} c_{\alpha}$                    & $- y_{\eta} s_{\theta} s_{\alpha} e^{i \tau}$ & $(1, 2)$ & $a c_{2 \theta} + \sqrt{2} v_H \left( \lambda_{H \phi} - \lambda_{H \eta} \right) s_{\theta} c_{\theta}$ \\
$(2, 1)$ & $- y_{\phi} s_{\theta} s_{\alpha} e^{- i \tau}$ & $y_{\eta} c_{\theta} c_{\alpha}$                  & $(2, 1)$ & $a c_{2 \theta} + \sqrt{2} v_H \left( \lambda_{H \phi} - \lambda_{H \eta} \right) s_{\theta} c_{\theta}$ \\
$(2, 2)$ & $y_{\phi} s_{\theta} c_{\alpha}$                    & $y_{\eta} c_{\theta} s_{\alpha} e^{i \tau}$   & $(2, 2)$ & $a s_{2 \theta} + \sqrt{2} v_H \left( \lambda_{H \phi} s_{\theta}^2 + \lambda_{H \eta} c_{\theta}^2 \right)$ \\ \hline
\end{tabular}
\caption{Yukawa couplings for the muon and exotic particles and scalar trilinear couplings in Eq.~\eqref{eq:relevantterms}.}
\label{tab:yLRAij}
\end{center}
\end{table}
These couplings lead to the radiative mass and effective Yukawa coupling of the muon at the one-loop level, through diagrams in Fig.~\ref{fig:MFdiagrams}:
\begin{align}
\mathcal{L}_{\rm eff} &\supset - m_{\mu}^{\rm rad} \bar{\mu}_L \mu_R - \frac{y_{\mu}^{\rm eff}}{\sqrt{2}} \bar{\mu}_L \mu_R h^0 + {\rm h.c.} \, , \\[1.2ex]
m_{\mu}^{\rm rad} &= \sum_{i, a} \frac{y_L^{i a} y_R^{i a}}{16 \pi^2} m_{\psi_a} \left( \frac{m_{\psi_a}^2 B_0 (0, m_{\psi_a}^2, m_{\psi_a}^2) - m_{\varphi_i^+}^2 B_0 (0, m_{\varphi_i^+}^2, m_{\varphi_i^+}^2)}{m_{\varphi_i^+}^2 - m_{\psi_a}^2} - 1 \right) \nonumber \\[0.6ex]
&= \frac{y_{\phi} y_{\eta}}{16 \pi^2} \frac{s_{2 \theta} s_{2 \alpha}}{4} \mathcal{F} (x_{1, 1}, x_{1, 2}, x_{2, 1}, x_{2, 2}) \, , \label{eq:radiativemmugene} \\[1.0ex]
y_{\mu}^{\rm eff} (p_{h^0}^2) &= - \sum_{i, j, a} \frac{y_L^{i a} y_R^{j a} A_{i j}}{16 \pi^2} m_{\psi_a} C_0 (m_{\mu}^2, m_{\mu}^2, p_{h^0}^2, m_{\varphi_i^+}^2, m_{\psi_a}^2, m_{\varphi_j^+}^2) \, , \label{eq:Yeffmu}
\end{align}
where $p_{h^0}$ is the four-momentum of the SM Higgs boson, $m_{\psi_a} \equiv \sqrt{m_{\psi_a}^2}$, and $B_0 (0, m^2, m^2)$ and $C_0 (m_{\mu}^2, m_{\mu}^2, p_{h^0}^2, m_{\varphi_i^+}^2, m_{\psi_a}^2, m_{\varphi_j^+}^2)$ denote loop integrals for the self-energy and triangle type diagrams, respectively, whose explicit forms are summarized in Appendix~\ref{app:LoopIntegrals}, and $\mathcal{F} (x_{1,1}, x_{1,2}, x_{2,1}, x_{2,2})$ is defined as
\begin{align}
\mathcal{F} (x_{1, 1}, x_{1, 2}, x_{2, 1}, x_{2, 2}) &\equiv - m_{\psi_1} e^{- i \tau} \left( \frac{x_{1, 1}}{1 - x_{1, 1}} \ln x_{1, 1} - \frac{x_{2, 1}}{1 - x_{2, 1}} \ln x_{2, 1} \right) \nonumber \\
&\hspace{4.0em} + m_{\psi_2} e^{i \tau} \left( \frac{x_{1, 2}}{1 - x_{1, 2}} \ln x_{1, 2} - \frac{x_{2, 2}}{1 - x_{2, 2}} \ln x_{2, 2} \right) \, , \label{eq:Ffunc}
\end{align}
with $x_{i,a} \equiv m_{\varphi_i^+}^2/m_{\psi_a}^2$.  Eq.~\eqref{eq:Yeffmu} neglects sub-dominant contributions with a chirality flip on the muon line. 
We show the full form for $y_{\mu}^{\rm eff}$ at the one-loop order in Appendix~\ref{app:FullForms}. 
To numerically evaluate the loop functions with a non-zero $p_{h^0}^2$, we use \texttt{LoopTools}~\cite{Hahn:1998yk}. 
Note that due to the radiatively generated muon mass, there is no standard relation between $m_{\mu}^{\rm rad}$ and $y_{\mu}^{\rm eff}$, namely, $m_{\mu}^{\rm rad} \neq y_{\mu}^{\rm eff} v_H / \sqrt{2}$. 
Hence, we need to check if the model satisfies a constraint from the measurement of $h \to \mu^+ \mu^-$. 
We discuss this constraint in Sec.~\ref{decaytomumu}. 
%%%%%%%%%%%%%%%%%%%%%%%%%%%%%%%%%
%%%%%%%%%%%%%%%%%%%%%%%%%%%%%%%%%

Since $m_{\mu}^{\rm rad}$ in Eq.~\eqref{eq:radiativemmugene} generally has a phase due to complex couplings $y_{L, R}^{i a}$, we need to remove it by a chiral rotation of the muon field as
\begin{align}
\mu \to e^{- i \theta_{\mu} \gamma_5 / 2} \mu \, ,
\label{eq:chiralrotmu}
\end{align}
where $\theta_{\mu}$ is defined as $m_{\mu}^{\rm rad} = m_{\mu} e^{i \theta_{\mu}}$. 
Here, $m_{\mu}$ is understood as the observed muon mass and a real value, and $\theta_{\mu}$ can be obtained as $\theta_{\mu} = \arg \bigl[ \mathcal{F} (x_{1, 1}, x_{1, 2}, x_{2, 1}, x_{2, 2})) \bigr]$. 
This rotation affects dipole operators,
\begin{align}
\mathcal{L}_{\rm dipole} = - \frac{e}{2} C_T (q^2) \left( \bar{\mu} \sigma^{\alpha \beta} \mu \right) F_{\alpha \beta} - \frac{e}{2} C_{T'} (q^2) \left( \bar{\mu} i \sigma^{\alpha \beta} \gamma_5 \mu \right) F_{\alpha \beta} \, ,
\label{eq:dipoleOPs}
\end{align}
where $q$ is the four-momentum for the photon. 
If there was no chiral rotation, $C_T (0)$ and $C_{T'} (0)$ would be the muon $g-2$ $a_{\mu}$ and the muon EDM $d_{\mu}$, respectively. 
After performing the chiral rotation of Eq.~\eqref{eq:chiralrotmu}, we can obtain the correct forms of $a_{\mu}$ and $d_{\mu}$ in our model as
\begin{align}
\mathcal{L}_{\rm dipole} &= - \frac{e}{4 m_{\mu}} a_{\mu} \left( \bar{\mu} \sigma^{\alpha \beta} \mu \right) F_{\alpha \beta} - \frac{i}{2} d_{\mu} \left( \bar{\mu} \sigma^{\alpha \beta} \gamma_5 \mu \right) F_{\alpha \beta} \, , \\[0.5ex]
a_{\mu} &= 2 m_{\mu} \left( C_T (0) \cos \theta_{\mu} + C_{T'} (0) \sin \theta_{\mu} \right) \, , \label{eq:amu} \\
d_{\mu} &= e \left( C_{T'} (0) \cos \theta_{\mu} - C_T (0) \sin \theta_{\mu} \right) \, . \label{eq:dmu}
\end{align}
The leading contributions to $C_T (0)$ and $C_{T'} (0)$ can be estimated from the diagram (b) in Fig.~\ref{fig:MFdiagrams} as
\begin{align}
C_T (0) &= \sum_{i, a} \frac{{\rm Re}[y_L^{i a} y_R^{i a}]}{16 \pi^2} m_{\psi_a} \Bigl[ C_0 (m_{\psi_a}^2, m_{\varphi_i^+}^2) + 2 C_1 (m_{\psi_a}^2, m_{\varphi_i^+}^2) \Bigr] \, , \label{eq:CT0} \\
C_{T'} (0) &= \sum_{i, a} \frac{{\rm Im}[y_L^{i a} y_R^{i a}]}{16 \pi^2} m_{\psi_a} \Bigl[ C_0 (m_{\psi_a}^2, m_{\varphi_i^+}^2) + 2 C_1 (m_{\psi_a}^2, m_{\varphi_i^+}^2) \Bigr] \, , \label{eq:CTp0}
\end{align}
where
\begin{align}
C_0 (m_{\psi_a}^2, m_{\varphi_i^+}^2) &\equiv C_0 (m_{\mu}^2, m_{\mu}^2, 0, m_{\varphi_i^+}^2, m_{\psi_a}^2, m_{\varphi_i^+}^2) \approx C_0 (0, 0, 0, m_{\varphi_i^+}^2, m_{\psi_a}^2, m_{\varphi_i^+}^2) \, , \\
C_1 (m_{\psi_a}^2, m_{\varphi_i^+}^2) &\equiv C_1 (m_{\mu}^2, 0, m_{\mu}^2, m_{\psi_a}^2, m_{\varphi_i^+}^2, m_{\varphi_i^+}^2)  \approx C_1 (0, 0, 0, m_{\psi_a}^2, m_{\varphi_i^+}^2, m_{\varphi_i^+}^2) \, ,
\end{align}
are loop integrals for the triangle type diagram, and approximations in the right hand sides are valid when $m_{\mu}^2 \ll m_{\varphi_i^+}^2$, $m_{\psi_a}^2$. 
In this case, we can obtain the following analytical forms of $C_0$ and $C_1$:
\begin{align}
C_0 (0, 0, 0, m_{\varphi_i^+}^2, m_{\psi_a}^2, m_{\varphi_i^+}^2) &= \frac{1}{m_{\psi_a}^2} \left[ \frac{1}{1 - x_{i, a}} + \frac{1}{(1 - x_{i, a})^2} \ln x_{i, a} \right] \, , \label{eq:C0approx} \\
C_1 (0, 0, 0, m_{\psi_a}^2, m_{\varphi_i^+}^2, m_{\varphi_i^+}^2) &= - \frac{1}{m_{\psi_a}^2} \left[ \frac{3 - x_{i, a}}{4 (1 - x_{i, a})^2} + \frac{1}{2 (1 - x_{i, a})^3} \ln x_{i, a} \right] \, . \label{eq:C1approx}
\end{align}
Here, $x_{i, a}$ is defined below Eq.~\eqref{eq:Ffunc}. 
Since the leading contributions to $C_T (0)$ and $C_{T'} (0)$ are the same except for the overall couplings, ${\rm Re}[y_L^{i a} y_R^{i a}]$ and ${\rm Im}[y_L^{i a} y_R^{i a}]$, we can expect a sufficiently large $d_{\mu}$ to be probed in near-future experiments when the muon $g-2$ is predicted to be $\mathcal{O}(10^{-9})$. 
That is, when $C_T (0) \cos \theta_{\mu} + C_{T'} (0) \sin \theta_{\mu} \sim C_{T'} (0) \cos \theta_{\mu} - C_T (0) \sin \theta_{\mu}$ is satisfied, we find
\begin{align}
d_{\mu} \sim \frac{e}{2 m_{\mu}} \times a_\mu
\simeq 2.34 \times 10^{-22} \, e \, {\rm cm} \, ,
\end{align}
with $a_\mu \simeq 2.51 \times 10^{-9}$. 
By using couplings in Table.~\ref{tab:yLRAij} and Eqs.~\eqref{eq:radiativemmugene}, \eqref{eq:C0approx} and \eqref{eq:C1approx}, $C_T (0)$ and $C_{T'} (0)$ can be rewritten as
\begin{align}
C_T (0) &= \frac{m_{\mu} {\cos \tau}}{\bigl| \mathcal{F} (x_{1, 1}, x_{1, 2}, x_{2, 1}, x_{2, 2}) \bigr|} \sum_{i, a = 1}^2 \frac{{(-1)^{i + a}}}{m_{\psi_a}} \frac{x_{i, a}^2 - 1 - 2 x_{i, a} \ln x_{i, a}}{2 (1 - x_{i, a})^3} \, , \label{eq:CT0exp} \\
C_{T'} (0) &= \frac{m_{\mu} {\sin \tau}}{\bigl| \mathcal{F} (x_{1, 1}, x_{1, 2}, x_{2, 1}, x_{2, 2}) \bigr|} \sum_{i, a = 1}^2 \frac{{(-1)^i}}{m_{\psi_a}} \frac{x_{i, a}^2 - 1 - 2 x_{i, a} \ln x_{i, a}}{2 (1 - x_{i, a})^3} \, . 
\label{eq:CTp0exp}
\end{align}
As $\bigl| \mathcal{F} \bigr|$ is proportional to $m_{\psi_{1,2}}$, their scalings are consistent with the rough estimation given in Eq.~\eqref{eq:roughEDMrad}. 
It is notable that when we change $\theta_{\rm phys} \to - \theta_{\rm phys}$, signs of $\sin \tau$ and $\sin \theta_{\mu}$ are flipped, the former of which leads to $C_{T'} (0) \to - C_{T'} (0)$ through Eq.~\eqref{eq:CTp0exp}, and hence, this change results in $d_{\mu} \to - d_{\mu}$ with $a_{\mu}$ unchanged. 
This fact tells us that it is enough to focus on the range $0 < \theta_{\rm phys} < \pi / 2$, because we are only interested in the prediction of $|d_{\mu}|$ here. 
Furthermore, $\theta_{\rm phys} = 0$ corresponds to a CP conserving limit which gives $|d_{\mu}| = 0$, while $\theta_{\rm phys} = \pi / 2$ leads to $\tau \approx \pi / 2$ unless $m_{LL} = m_{RR}$ (see Eq.~\eqref{eq:tantau}), predicting $\cos \theta_{\mu} \approx 0$, and hence, $|d_{\mu}| \propto C_{T'} (0) \cos \theta_{\mu} - C_T (0) \sin \theta_{\mu} \approx 0$. 
Hereafter, we denote $d_{\mu}$ as its absolute value $|d_{\mu}|$ in our analysis.

%#######################
\subsection{Muon coupling constraints}
\label{decaytomumu}

In our model, the muon Yukawa coupling to the Higgs boson is generated at the one-loop level and does not follow the standard relation, $m_{\mu}^{\rm rad} \neq y_{\mu}^{\rm eff} v_H / \sqrt{2}$. 
The ATLAS~\cite{ATLAS:2020fzp} and CMS~\cite{CMS:2020xwi} experiments have searched for the Higgs boson decay $h \to \mu^+ \mu^-$, which lead to constraints on the $h$-$\mu$-$\mu$ coupling as
\begin{align}
\left| \kappa_{\mu} \right| &< 1.47 \quad ({\rm ATLAS}) \, , \label{eq:constATLAS} \\[1ex]
0.61 < \left| \kappa_{\mu} \right| &< 1.44 \quad ({\rm CMS}) \, , \label{eq:constCMS}
\end{align}
where we use BR$(h \to \mu^+ \mu^-)_{\rm SM} \simeq 2.16 \times 10^{-4}$ for $m_h = 125.25$ GeV~\cite{LHCHiggsCrossSectionWorkingGroup:2016ypw}, and $\kappa_{\mu}$ is defined by comparing the decay width of $h \to \mu^+ \mu^-$ to that of the SM,
\begin{align}
\Gamma_{h \to \mu^+ \mu^-}^{\rm SM} = \frac{m_h}{8 \pi} \left( \frac{m_{\mu}}{v_H} \right)^2 \left( 1 - \frac{4 m_{\mu}^2}{m_h^2} \right)^{3/2} \, .
\end{align}
In our model, the width of $h \to \mu^+ \mu^-$ is estimated as
\begin{align}
\Gamma_{h \to \mu^+ \mu^-} = \frac{m_h}{16 \pi} \sqrt{1 - \frac{4 m_{\mu}^2}{m_h^2}} \left[ \left( 1 - \frac{4 m_{\mu}^2}{m_h^2} \right) \left( {\rm Re} \: y_{\mu}^{\rm eff} \right)^2 + \left( {\rm Im} \: y_{\mu}^{\rm eff} \right)^2 \right] \, .
\end{align}
Then, we find
\begin{align}
\left| \kappa_{\mu} \right| &= \frac{1}{\sqrt{2}} \frac{v_H}{m_{\mu}} \sqrt{\left( {\rm Re} \: y_{\mu}^{\rm eff} \right)^2 + \left( 1 - \frac{4 m_{\mu}^2}{m_h^2} \right)^{-1} \left( {\rm Im} \: y_{\mu}^{\rm eff} \right)^2} \approx \frac{1}{\sqrt{2}} \frac{v_H}{m_{\mu}} \left| y_{\mu}^{\rm eff} \right| \, .
\label{eq:kappamu}
\end{align}
Here, we have used $4 m_{\mu}^2 \ll m_h^2$. 

Since exotic particles exclusively couple to the muon, the ratio between the $Z \to e^+ e^-$ and $Z \to \mu^+ \mu^-$ decay widths may constrain our parameter space. 
The current experimental status for this ratio is
\cite{ALEPH:2005ab}
\begin{align}
\frac{\Gamma (Z \to \mu^+ \mu^-)}{\Gamma (Z \to e^+ e^-)} = 1.0009 \pm 0.0028 \, .
\label{eq:LFUZemu}
\end{align}
The muon couplings to the $Z$ boson can be parameterized as
\begin{align}
\mathcal{L}_Z \supset \frac{g}{\cos \theta_W} \bar{\mu} \gamma^{\alpha} \Bigl[ (g_L^{\mu} + \delta g_L^{\mu}) P_L + (g_R^{\mu} + \delta g_R^{\mu}) P_R \Bigr] \mu Z_{\alpha} \, ,
\end{align}
where $g$ denotes the $SU(2)_L$ gauge coupling, $\theta_W$ is the weak mixing angle, and $g_L^{\mu} = - \frac{1}{2} + \sin^2 \theta_W$, and $g_R^{\mu} = \sin^2 \theta_W$ are the muon couplings to the $Z$ boson in the SM. 
In our model, new physics contributions $\delta g_{L,R}$ are induced by the diagram (b) in Fig.~\ref{fig:MFdiagrams} with replacing the photon to the $Z$ boson, and their expressions are found in ref.~\cite{Baker:2020vkh}. 
The ratio in Eq.~\eqref{eq:LFUZemu} is then estimated as
\begin{align}
\frac{\Gamma (Z \to \mu^+ \mu^-)}{\Gamma (Z \to e^+ e^-)} \simeq 1 + \frac{2 g_L^e {\rm Re} \left( \delta g_L^{\mu} \right) + 2 g_R^e {\rm Re} \left( \delta g_R^{\mu} \right)}{(g_L^e)^2 + (g_R^e)^2} \equiv 1 + \delta_{\mu \mu} \, , \label{eq:LFUZnewphysics}
\end{align}
where $g_{L,R}^e = g_{L,R}^{\mu}$ are the electron couplings to the $Z$ boson in the SM, and we assume that new physics contributions are smaller than those of the SM, $\delta g_{L,R}^{\mu} \ll g_{L,R}^{\mu}$. 
Then, Eq.~\eqref{eq:LFUZemu} indicates that $|\delta_{\mu \mu}|$ must be less than $\mathcal{O}(10^{-3})$.

%#######################
\section{Dark matter}
\label{DMpheno}

The candidate of DM in our model is the lightest Majorana fermion $\psi_1^M$ or the lightest neutral scalar $a_{\phi}$, depending on their masses. 
In the present paper, we focus on the case that $\psi_1^M$ is the lightest exotic particle, and hence, gives the DM candidate. 
Hereafter, we denote $\psi_1^M$ as $\psi_1$ for simplicity. 
For the case that $a_{\phi}$ is the DM candidate, there is no direct correlation with the muon EDM, because the mass $m_{a_\phi}$ does not contribute to the muon EDM at the one-loop level. 

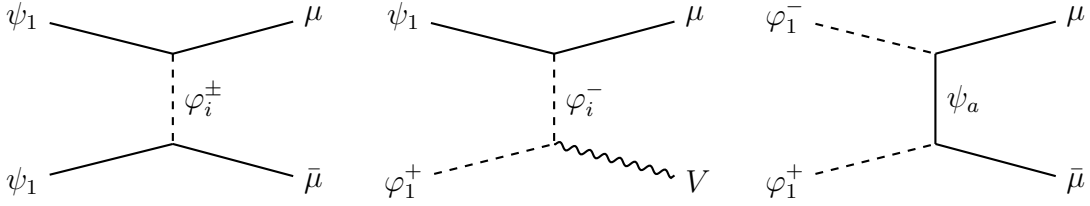
\begin{figure}[t]
\begin{center}
\begin{tikzpicture}
\begin{feynman}[large]
%%% DM self-annihilation %%%
\vertex (a1) {\(\psi_1\)};
\vertex [right=2.0cm of a1] (b1);
\vertex [below=0.5cm of b1] (c1);
\vertex [right=1.6cm of b1] (d1) {\(\mu\)};
\vertex [below=1.2cm of c1] (e1);
\vertex [below=0.5cm of e1] (f1);
\vertex [below=2.2cm of a1] (g1) {\(\psi_1\)};
\vertex [right=1.6cm of f1] (h1) {\(\bar{\mu}\)};
\diagram [medium] {
(a1) -- (c1) -- (d1),
(c1) -- [scalar, edge label=\(\varphi_i^{\pm}\)] (e1),
(g1) -- (e1) -- (h1),
};
%%% DM-charged scalar coannihilation %%%
\vertex [right=1.2cm of d1] (a2) {\(\psi_1\)};
\vertex [right=2.0cm of a2] (b2);
\vertex [below=0.5cm of b2] (c2);
\vertex [right=1.6cm of b2] (d2) {\(\mu\)};
\vertex [below=1.2cm of c2] (e2);
\vertex [below=0.5cm of e2] (f2);
\vertex [below=2.2cm of a2] (g2) {\(\varphi_1^+\)};
\vertex [right=1.6cm of f2] (h2) {\(V\)};
\diagram [medium] {
(a2) -- (c2) -- (d2),
(c2) -- [scalar, edge label=\(\varphi_i^-\)] (e2),
(g2) -- [scalar] (e2) -- [photon] (h2),
};
%%% charged scalar self-annihilation %%%
\vertex [right=1.2cm of d2] (a3) {\(\varphi_1^-\)};
\vertex [right=2.0cm of a3] (b3);
\vertex [below=0.5cm of b3] (c3);
\vertex [right=1.6cm of b3] (d3) {\(\mu\)};
\vertex [below=1.2cm of c3] (e3);
\vertex [below=0.5cm of e3] (f3);
\vertex [below=2.2cm of a3] (g3) {\(\varphi_1^+\)};
\vertex [right=1.6cm of f3] (h3) {\(\bar{\mu}\)};
\diagram [medium] {
(a3) -- [scalar] (c3) -- (d3),
(c3) -- [edge label=\(\psi_a\)] (e3),
(g3) -- [scalar] (e3) -- (h3),
};
\end{feynman}
\end{tikzpicture}
\end{center}
\vspace{-0.5cm}
\caption{Feynman diagrams for DM self-annihilation (left), DM-charged scalar coannihilation (center) and charged scalar self-annihilation (right). 
$V$ denotes the $\gamma/Z$ boson. 
Other diagrams can be obtained by changing final states with an appropriate mediator.}
\label{fig:DMannihilation}
\end{figure}

The main annihilation mode of the DM fermion is $\psi_1 \psi_1 \to \mu \bar{\mu}$ through the $t$-channel exchange of $\varphi_i^\pm$, as shown in the left of Fig.~\ref{fig:DMannihilation}. 
%%%%%%%%%%%%%%%%%%%%%%%%%%%%%%%%%
%%%%%%%%%%%%%%%%%%%%%%%%%%%%%%%%%
In the expansion of the thermally averaged cross section by the DM velocity $v$, $\langle \sigma v \rangle_{\mu \bar{\mu}} = a_{\mu \bar{\mu}} + b_{\mu \bar{\mu}} v^2 + \mathcal{O}(v^4)$, $s$-wave and $p$-wave contributions are given by
\begin{align}
a_{\mu \bar{\mu}} &= \frac{1}{16 \pi m_{\psi_1}^2} \left[ \left| \sum_{i = 1,2} \frac{y_L^{i 1} y_R^{i 1}}{1 + x_{i, 1}} \right|^2 + \mathcal{O}\left( \frac{m_{\mu}^2}{m_{\psi_1}^2} \right) \right] \, , \label{eq:sigvmumuswave} \\
b_{\mu \bar{\mu}} &= \frac{1}{48 \pi m_{\psi_1}^2} \left( \sum_{i = 1,2} \frac{|y_L^{i 1}|^2}{(1 + x_{i, 1})^2} \sqrt{1 + x_{i, 1}^2} \right)^2 + (L \to R) \, , \label{eq:sigvmumupwave}
\end{align}
where the second term in Eq.~\eqref{eq:sigvmumuswave} is suppressed by $m_\mu/m_{\psi_1}$. 
Thus, the $s$-wave contribution dominates the total DM annihilation cross section in our focused parameter space. 
Note that for the annihilation mode $\psi_1 \psi_1 \to \nu_{\mu} \bar{\nu}_{\mu}$, the $s$-wave contribution is suppressed by a tiny neutrino mass, because there is no right-handed coupling $y_R^{i 1}$ for the neutrino. 
The other annihilation cross sections, such as $\psi_1 \psi_1 \to \gamma \gamma$ and $\psi_1 \psi_1 \to \mu \bar{\mu} \gamma$, are several orders of magnitude smaller than that of $\psi_1 \psi_1 \to \mu \bar{\mu}$. 

In the thermal freeze-out scenario, the number density of DM is calculated by the Boltzmann equation,
\begin{align}
\frac{d n_{\psi_1}(t)}{d t} + 3 H(t) n_{\psi_1}(t) = - \langle \sigma v \rangle_{\rm eff} \Bigl[ n_{\psi_1}(t)^2 - n_{\psi_1}^{\rm eq}(t)^2 \Bigr] \, .
\label{eq:BoltzmannEq}
\end{align}
Here, $H(t)$ denotes the Hubble rate, and $n_{\psi_1}$ is the number density of $\psi_1$, while $n_{\psi_1}^{\rm eq}$ is that in equilibrium. 
The effective annihilation cross section $\langle \sigma v \rangle_{\rm eff}$ is estimated by summing all possible annihilation modes, i.e., $\psi_1 \psi_1 \to \ell \bar{\ell}, V V', \ell \bar{\ell} V$ ($\ell = \mu, \nu; V, V' = \gamma, Z, W$). 
However, when the DM and charged scalar masses are almost degenerate, coannihilation processes should be taken into account for solving the Boltzmann equation. 
In this case, we have~\cite{Griest:1990kh}
\begin{align}
(\sigma v)_{\rm eff} &= \frac{1}{(g_{\psi_1} + \bar{g}_{\varphi_1^+})^2} \Bigl[ g_{\psi_1}^2 (\sigma v)_{\psi_1 \psi_1} + g_{\psi_1} \bar{g}_{\varphi_1^+} (\sigma v)_{\psi_1 \varphi_1^+} + \bar{g}_{\varphi_1^+}^2 (\sigma v)_{\varphi_1^+ \varphi_1^-} \Bigr] \, , \label{eq:sigveffcoannihi} \\[0.5ex]
&\text{with } \bar{g}_{\varphi_1^+} = g_{\varphi_1^+} \left( \frac{m_{\varphi_1^+}}{m_{\psi_1}} \right)^{3/2} e^{- (m_{\varphi_1^+} - m_{\psi_1})/T} \, , \label{eq:gbarcoannihi}
\end{align}
where $g_{\psi_1} = 2$ and $g_{\varphi_1^+} = 2$ are internal degrees of freedom for $\psi_1$ and $\varphi_1^+$, respectively, $T$ is the temperature, and $(\sigma v)_{XY}$ denotes the (co)annihilation cross section whose initial state is $XY$. 
The corresponding diagrams are shown in Fig.~\ref{fig:DMannihilation}. 
The second term in Eq.~\eqref{eq:sigveffcoannihi} is suppressed by the exponential factor in Eq.~\eqref{eq:gbarcoannihi} and the third term is more suppressed due to the squared exponential factor when $m_{\varphi_1^+} \gg m_{\psi_1}$. 
As $m_{\varphi_1^+}$ decreases and is close to $m_{\psi_1}$, the second term gives a non-negligible contribution to $(\sigma v)_{\rm eff}$~\cite{Gondolo:1990dk,Edsjo:1997bg}. 
The resultant DM relic density is given by
\begin{align}
\Omega h^2 = \frac{m_{\psi_1} n_{\psi_1}(t_0)}{\rho_c} h^2 \, ,
\end{align}
where $\rho_c$ is the critical density of the Universe and $n_{\psi_1}(t_0)$ is the today's number density of $\psi_1$ obtained by solving the Boltzmann equation~\eqref{eq:BoltzmannEq}. 
To calculate the DM relic density $\Omega h^2$ including appropriate coannihilation processes, we use \texttt{micrOMEGAs\_5.2.13}~\cite{Belanger:2018ccd,Belanger:2020gnr}. 

Although our DM particle $\psi_1$ does not couple to the SM quarks and gluons, the DM-nucleon scattering is induced by contact and non-contact type interactions. 
In our model, relevant interactions for the scattering are
\begin{align}
\mathcal{L}_{\rm eff} \supset a_{\psi_1} \overline{\psi_1} \gamma^{\mu} \gamma^5 \psi_1 \partial^{\nu} F_{\mu \nu} + C_{S,p} \overline{\psi_1} \psi_1 \bar{p} p + C_{S,n} \overline{\psi_1} \psi_1 \bar{n} n \, .
\label{eq:Leff-DD}
\end{align}
Here, $p$ and $n$ represent the proton and the neutron, and the effective coefficients are estimated as
\begin{align}
a_{\psi_1} &= - \frac{e}{16 \pi^2 m_{\psi_1}^2} \sum_{i = 1, 2} \Bigl( |y_L^{i 1}|^2 + |y_R^{i 1}|^2 \Bigr) \hat{a}_{\psi_1} (x_{i,1}, \epsilon_{\mu}) \, , \\[0.5ex]
C_{S,N} &= - m_N \sum_q C_{S,q} \frac{f_{T q}^{(N)}}{m_q} = - \frac{y_{\psi_1}^{\rm eff} m_N}{\sqrt{2} m_h^2 v_H} \sum_q f_{T q}^{(N)} \qquad (N = p, n) \, ,
\end{align}
where $x_{i, 1}$ is defined below Eq.~\eqref{eq:Ffunc}, $\epsilon_{\mu} \equiv m_{\mu}^2 / m_{\psi_1}^2$, $\hat{a}_{\psi_1} (x, y)$ is the loop function for the anapole operator, which is given by
\begin{align}
\hat{a}_{\psi_1} (x, y) = \frac{1}{12} \left[ \frac{3}{2} \ln \left( \frac{y}{x} \right) + \frac{3 x - 3 y + 1}{\Delta (x, y)^{1/2}} \tanh^{-1} \left( \frac{\Delta (x, y)^{1/2}}{x + y - 1} \right) \right] \, ,
\end{align}
with $\Delta (x, y) = x^2 + (y - 1)^2 - 2 x (y + 1)$, $C_{S,q}$ denotes the effective coupling of an operator $\overline{\psi_1} \psi_1 \bar{q} q$ with the SM quark $q$, and $f_{T q}^{(N)}$ is related to the quark mass contribution to the nucleon mass, whose value can be found in refs.~\cite{DelNobile:2021wmp,Ellis:2000ds,Gondolo:2004sc,Ellis:2008hf,Belanger:2008sj,Cheng:2012qr,Belanger:2013oya,Crivellin:2013ipa,Hoferichter:2015dsa,Ellis:2018dmb}. 
$y_{\psi_1}^{\rm eff}$ is the effective Yukawa coupling of $\psi_1$, and it can be obtained by the replacement of $m_{\psi_1} \leftrightarrow m_{\mu}$ and $y_L \leftrightarrow y_R$ in the expression of $y_{\mu}^{\rm eff}$ given in Eq.~\eqref{eq:YeffFull}. 
This effective Yukawa coupling increases when $m_{\psi_1}$ becomes large, because it is proportional to $m_{\psi_1}$ like $y_{\mu}^{\rm eff}$ (see Eq.~\eqref{eq:Yeffmu}). 
The function $\hat{a}_{\psi_1} (x, y)$ is enhanced when $x \to 1$ with $y \neq 0$. 
Therefore, the limit of $m_{\psi_1} \simeq m_{\varphi_1^+}$ leads to a large contribution to the cross section from $a_{\psi_1}$. 
Note that for the Majorana DM model, there are other contributions through the $Z$-penguin which lead to effective interactions such as $(\overline{\psi_1} \gamma^{\mu} \gamma^5 \psi_1) (\bar{q} \gamma_{\mu} q)$ and $(\overline{\psi_1} \gamma^{\mu} \gamma^5 \psi_1) (\bar{q} \gamma_{\mu} \gamma^5 q)$. 
However, these contributions are suppressed by the lepton mass (and the DM velocity for the former interaction), and we neglect their effects in our analysis. 
Using the effective couplings in Eq.~\eqref{eq:Leff-DD}, the differential cross section with respect to the recoil energy $E_R$ is estimated as
\begin{align}
\frac{d \sigma}{d E_R} &= \left\{ \frac{2 m_N f_A^2}{\pi v^2} + \frac{4 \alpha Z^2}{v^2} a_{\psi_1}^2 \left[ 2 m_N v^2 - \frac{(m_N + m_{\psi_1})^2}{m_{\psi_1}^2} E_R \right] \right\} | F_{\rm Helm} (E_R) |^2 \nonumber \\[0.5ex]
&\hspace{1.2em}+ \frac{8 m_N^2 \mu_A^2}{\pi v^2} E_R a_{\psi_1}^2 \frac{J_A + 1}{3 J_A} | F_{\rm spin} (E_R) |^2 \, ,
\label{eq:dsigdER}
\end{align}
where $v$ is the DM velocity, $\alpha$ is the fine structure constant, $f_A = Z C_{S,p} + (A - Z) C_{S,n}$ with an atomic number $Z$ and a mass number $A$, and $m_N$, $\mu_A$ and $J_A$ are the mass, magnetic moment and spin of the nucleus, respectively. 
$F_{\rm Helm} (E_R)$ and $F_{\rm spin} (E_R)$ denote form factors found in refs.~\cite{Helm:1956zz,Lewin:1995rx}. 
It can be seen from Eq.~\eqref{eq:dsigdER} that the anapole contribution is suppressed by the DM velocity $v$ or the recoil energy $E_R$. 
On the other hand, there is no suppression for contributions from the contact-type interactions. 
It is notable that in our model, $C_{S,N}$ in $f_A$ is enhanced by $y_{\psi_1}^{\rm eff}$ due to the absence of the tree-level muon Yukawa coupling. 
Recently, the LUX-ZEPLIN (LZ) experiment has reported their first results for spin-independent (SI) and spin-dependent (SD) DM-nucleon scattering cross sections~\cite{LZ:2022ufs}. 
The upper limit on the SI cross section has been improved, compared with previous results from the XENON1T~\cite{XENON:2018voc,XENON:2019rxp} and PandaX-4T~\cite{PandaX-4T:2021bab,Liu:2022zgu} experiments. 
The corresponding cross sections in our model are given by~\cite{DelNobile:2021wmp,Pospelov:2000bq,Ho:2012bg}
\begin{align}
\sigma_{\rm SI}^{\rm scalar} &= \frac{4 \mu_N^2}{\pi} \frac{|y_{\psi_1}^{\rm eff}|^2}{2 m_h^4} \frac{m_N^2}{v_H^2} \left( \sum_q f_{T q}^{(N)} \right)^2 \, , \label{eq:sigSIscl} \\
\sigma_{\rm SI}^{\rm anapole} &= 8 \alpha |a_{\psi_1}|^2 \mu_N^2 v^2 \, . \label{eq:sigSIapl}
\end{align}
Here, $\mu_N$ is the reduced mass for $m_{\psi_1}$ and the nucleon mass $\simeq 0.939$ GeV.

%#######################
\section{Numerical analysis}
\label{results}

In this section, we first summarize the independent parameters in our model. 
Then, the parameter space that gives the correct DM relic density and explains the muon $g - 2$ anomaly is identified and the size of the muon EDM is indicated in that region as well as more general parameter regions. 
We take account of muon coupling constraints presented in section~\ref{decaytomumu} and also discuss constraints from DM direct and indirect detection experiments as well as collider searches.

%#######################
\subsection{Independent parameters}
\label{indepparams}

The Lagrangian of our model contains 18 parameters,
\begin{equation}
\begin{split}
&y_{\phi}, y_{\eta}, |m_D|, |m_{LL}|, |m_{RR}|, \theta_{\rm phys},  \\
&m_H^2, m_{\phi}^2, m_{\eta}^2, a, \lambda_H, \lambda_{\phi}, \lambda_{\eta}, \lambda_{H \phi}, \lambda_{H \eta}, \lambda_{\phi \eta}, \lambda'_{H \phi}, \lambda''_{H \phi} \, .
\end{split}
\end{equation}
Note that some of them are irrelevant to our analysis on the calculation of the muon $g-2$, the muon EDM, the radiative mass, and the effective Yukawa coupling of the muon. 
The Higgs mass-squared parameter $m_H^2$ is fixed by the minimization condition in Eq.~\eqref{eq:mincond}, and $\lambda_H$ should be determined so that the SM Higgs mass, $m_h = 125.25$ GeV, is correctly reproduced. 
The quartic couplings $\lambda_{\phi}$, $\lambda_{\eta}$ and $\lambda_{\phi \eta}$ are irrelevant to the mass spectrum of exotic particles, although these values should be consistent with perturbative unitarity bounds (commented below) and also chosen to avoid an unstable minimum of the scalar potential. 
Moreover, $y_{\phi} y_{\eta}$ can be fixed by using Eq.~\eqref{eq:radiativemmugene}, but we need to check that values of the couplings do not exceed $\sqrt{4 \pi}$. 
As a result, the relevant (and independent) input parameters for the analysis can be read as
\begin{align}
y_{\phi}, |m_D|, |m_{LL}|, |m_{RR}|, \theta_{\rm phys}, m_{\phi}^2, m_{\eta}^2, a, \lambda_{H \phi}, \lambda'_{H \phi}, \lambda''_{H \phi}, \lambda_{H \eta} \, .
\label{eq:indepparaMF}
\end{align}
Note that $\lambda'_{H \phi}$ and $\lambda''_{H \phi}$ are relevant only to the masses of heavy neutral scalars, $m_{\sigma_{\phi}}^2$ and $m_{a_{\phi}}^2$ (see Eq.~\eqref{eq:M20}), and irrelevant to our following analysis as long as the DM candidate of the model is $\psi_1$. 
Furthermore, we discuss our results by using $M_{\phi}^2$ and $M_{\eta}^2$ instead of $m_{\phi}^2, m_{\eta}^2, \lambda_{H \phi}$ and $\lambda_{H \eta}$ (see below Eq.~\eqref{eq:M20}). 

We here comment on perturbative unitarity bounds~\cite{Lee:1977yc,Lee:1977eg}, which are related to $2 \to 2$ scattering processes of scalar particles. 
At the tree level, it is clear that quartic couplings are related to their amplitudes. 
In addition, trilinear couplings also contribute to them through $s$-, $t$- and $u$-channel processes if scalar particles are not so heavy. 
There are studies on the bounds, e.g., for models extended by singlet scalars~\cite{Cynolter:2004cq,Kang:2013zba,Costa:2014qga} and doublet scalars~\cite{Casalbuoni:1986hy,Casalbuoni:1987eg,Maalampi:1991fb,Kanemura:1993hm,Ginzburg:2003fe,Akeroyd:2000wc,Horejsi:2005da}. 
Since our model is a hybrid extension with one singlet and one doublet scalars, there are lots of scattering processes like $h h \to \sigma_{\phi} \sigma_{\phi}$, $h \sigma_{\phi} \to \varphi_i^+ \varphi_j^-$ and $\varphi_i^+ \varphi_j^- \to a_{\phi} a_{\phi}$. 
To obtain perturbative unitarity bounds in our model, we use the \texttt{SARAH/SPheno} framework~\cite{Porod:2003um,Porod:2011nf,Staub:2008uz,Staub:2009bi,Staub:2010jh,Staub:2012pb,Staub:2013tta}. 
The details of the calculation for general scalar couplings can be found in ref.~\cite{Goodsell:2018tti}.

%#######################
\subsection{Results}
\label{sec:results}

%%%%%%%%%%%%%%%%%%%%%%%%%%%%%%%%%
\begin{figure}[t]
\centering
\includegraphics[width=0.65\textwidth,bb=0 0 400 418]{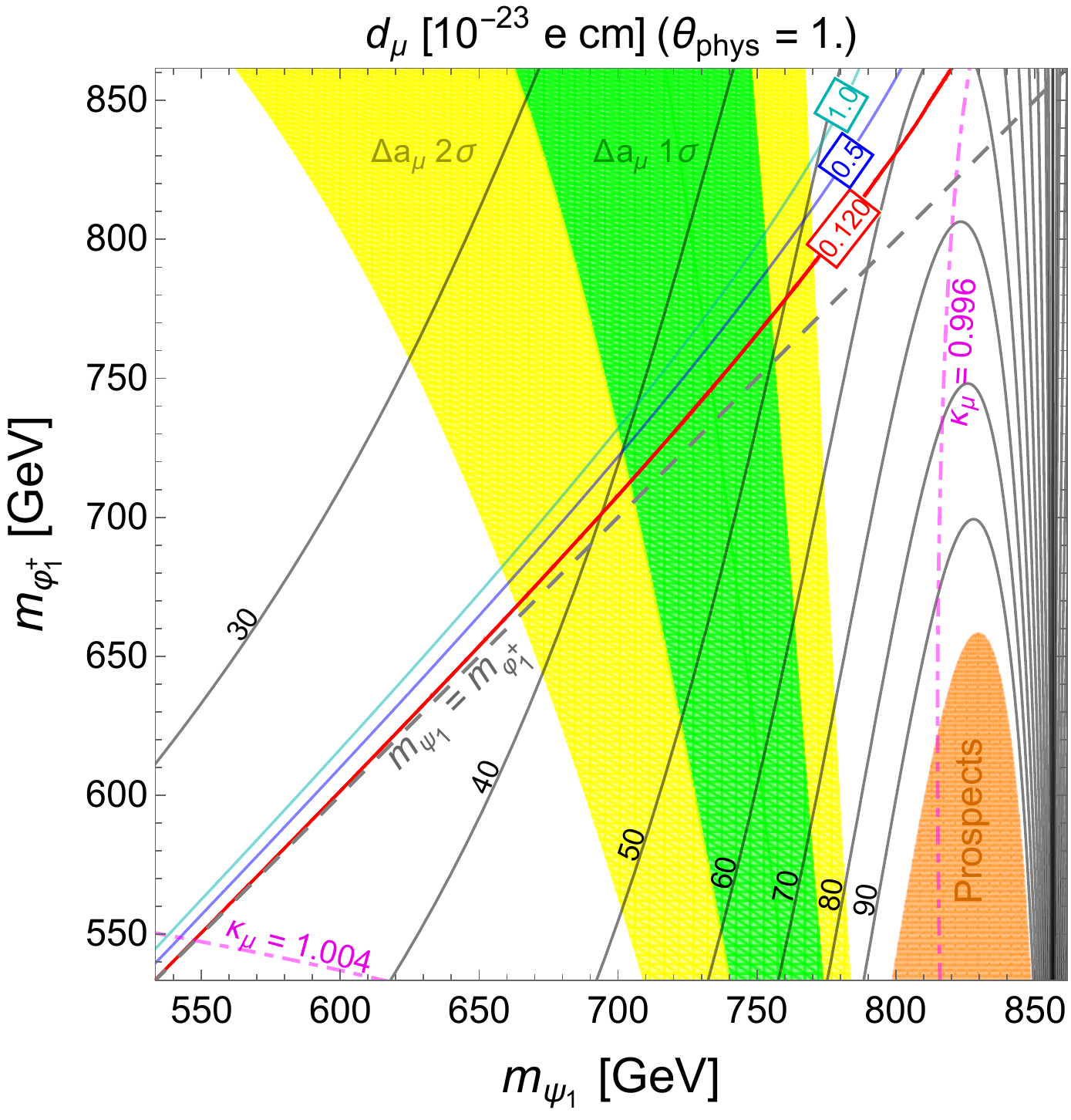}
\caption{The predictions of $(g-2)_{\mu}$ and the muon EDM $d_{\mu}$ in terms of $m_{\psi_1}$ and $m_{\varphi_1^+}$. 
We fix the model parameters as follows: $y_{\phi} = 1.2$, $m_D = 700 \, {\rm GeV}$, $m_{RR} = 1000 \, {\rm GeV}$, $\theta_{\rm phys} = 1.0$, $M_{\phi}^2 = (1000 \, {\rm GeV})^2$, $a = 900 \, {\rm GeV}$. $m_{LL}$ and $M_{\eta}^2$ are changed as $320 \, {\rm GeV} \leq m_{LL} \leq 1200 \, {\rm GeV}$ and $(540 \, {\rm GeV})^2 \leq M_{\eta}^2 \leq (1000 \, {\rm GeV})^2$, respectively. 
Green and yellow shaded regions denote $1\sigma$ and $2\sigma$ bands for $(g-2)_{\mu}$, and black lines correspond to contours for $d_{\mu}$ in $10^{-23} \, e \, {\rm cm}$ unit. 
The future prospects of the Fermilab Muon $g-2$~\cite{Chislett:2016jau} and J-PARC Muon $g-2$/EDM~\cite{Abe:2019thb} experiments are shown as the orange shaded region. 
The red band is the correct DM relic density, $\Omega h^2 = 0.120 \pm 0.001$, and contours for $\Omega h^2 = 0.5$ and $1.0$ are shown in blue and turquoise lines, respectively. 
The dot-dashed magenta lines correspond to $\kappa_{\mu} = 0.996$ and 1.004. 
The dashed gray line gives $m_{\psi_1} = m_{\varphi_1^+}$.}
\label{fig:resMFmodel}
\end{figure}
%%%%%%%%%%%%%%%%%%%%%%%%%%%%%%%%%
Fig.~\ref{fig:resMFmodel} shows the predictions of $(g-2)_{\mu}$ and the muon EDM $d_{\mu}$ in our model. 
Here, we fix the relevant parameters as~\footnote{If we change $m_{RR}$ instead of $m_{LL}$ and set $m_{LL} = 1000$ GeV, the prediction for $d_{\mu}$ is totally the same as that of Fig.~\ref{fig:resMFmodel}, because the mass eigenvalues of $\psi_{1,2}$ and the mixing angle $\alpha$ are symmetric under $m_{LL} \leftrightarrow m_{RR}$. 
Although $\tan \tau$ changes its sign, it is irrelevant to the absolute value of $d_{\mu}$.}
\begin{equation}
\begin{split}
&y_{\phi} = 1.2, \quad M_{\phi}^2 = (1000 \, {\rm GeV})^2, \quad a = 900 \, {\rm GeV},  \\
&m_D = 700 \, {\rm GeV}, \quad m_{RR} = 1000 \, {\rm GeV}, \quad \theta_{\rm phys} = 1.0,
\label{eq:benchmark}
\end{split}
\end{equation}
and change $m_{LL}$ and $M_{\eta}^2$ as $320 \, {\rm GeV} \leq m_{LL} \leq 1200 \, {\rm GeV}$ and $(540 \, {\rm GeV})^2 \leq M_{\eta}^2 \leq (1000 \, {\rm GeV})^2$, respectively. 
For this parameter choice, the lightest particle among $X$-odd particles is either $\psi_1$ or $\varphi_1^+$. 
The other parameters which are not shown in Eq.~\eqref{eq:indepparaMF}, i.e. scalar quartic couplings, does not affect the analysis here and are taken to be moderate values to satisfy perturbative unitarity bounds. 
Note that when the scalar trilinear coupling $a$ becomes large, some of the quartic couplings should be $\mathcal{O}(1)$ to avoid the instability of the vacuum to give the correct electroweak symmetry breaking. 
We numerically check that the SM vacuum is stable when all of quartic couplings are within the range of $0.2$-$0.5$ with the parameter choice shown above. 
These values of quartic couplings also satisfy perturbative unitarity bounds, which is checked by the \texttt{SARAH/SPheno} framework. 
In addition, since we have an additional $SU(2)_L$ doublet scalar, there is a new contribution to the $T$-parameter~\cite{Peskin:1990zt,Peskin:1991sw}. 
We have calculated the contribution by following refs.~\cite{Grimus:2007if,Grimus:2008nb} and found that our parameter choice leads to $\Delta T \sim 0.002$, which satisfies the current constraint~\cite{ParticleDataGroup:2022pth}. 

The current discrepancy of $(g-2)_{\mu}$ is~\cite{Muong-2:2006rrc,Keshavarzi:2018mgv,Aoyama:2020ynm,Muong-2:2021ojo}
\begin{align}
\Delta a_{\mu} = (2.51 \pm 0.59) \times 10^{-9} \, ,
\end{align}
whose $1\sigma$ and $2\sigma$ bands are shown as green and yellow shaded regions in Fig.~\ref{fig:resMFmodel}. 
Note that a lighter $m_{\psi_1}$ predicts a larger $\Delta a_{\mu}$ due to its dependence, $\Delta a_{\mu} \sim 1 / m_{\psi_1}^2$. 
In the figure, black lines correspond to contours for $d_{\mu}$ in $10^{-23} \, e \, {\rm cm}$ unit. 
The future prospect of the muon EDM, which is reported as $\mathcal{O}(10^{-21}) \, e \, {\rm cm}$ at the Fermilab Muon $g-2$ experiment~\cite{Chislett:2016jau} and the J-PARC Muon $g-2$/EDM experiment~\cite{Abe:2019thb}, is shown as the orange shaded region. 
The red band shows the parameter space where the correct DM relic density, $\Omega h^2 = 0.120 \pm 0.001$~\cite{Planck:2018vyg}, is obtained. 
Outside of this band, the relic density changes rapidly, as one can see from the blue and turquoise contours which correspond to $\Omega h^2 = 0.5$ and $1.0$, respectively. 
Note that in the whole parameter space of Fig.~\ref{fig:resMFmodel}, a new physics contribution to the ratio between the decay widths of $Z \to e^+ e^-$ and $Z \to \mu^+ \mu^-$ in Eq.~\eqref{eq:LFUZnewphysics} is sufficiently small, and we obtain $|\delta_{\mu \mu}| \lesssim 4 \times 10^{-4}$ which is consistent with the current data~\eqref{eq:LFUZemu}. 

For the case of $m_{\psi_1} > m_{\varphi_1^+}$ (the region below the dashed gray line), the $\psi_1 \to \varphi_1^{\pm} + \mu^{\mp}$ decay occurs at the tree level, and therefore, $\psi_1$ cannot be a DM candidate~\footnote{Since $m_{\psi_1} < m_{\sigma_{\phi}, a_{\phi}}$ in the current input parameters (see Eqs.~\eqref{eq:M20} and \eqref{eq:benchmark}), the DM cannot decay into $\nu_{\mu} + \sigma_{\phi}$, $\nu_{\mu} + a_{\phi}$ in the plotted region of Fig.~\ref{fig:resMFmodel}.}. 
Without any interaction to break the exotic number symmetry $X$, $\varphi_1^+$ is a stable exotic particle, which may be cosmologically dangerous. 
However, we can consider, for example, an interaction with the right-handed electron, $\overline{L_L^{\mu}} \phi^{\dagger} e_R$, to make $\varphi_1^+$ decay into $\nu_{\mu} + e^+$~\footnote{This lepton flavor violating (LFV) interaction does not induce LFV processes such as $\mu \to e \gamma$ because of the muon number symmetry $L_{\mu}$. 
However, the interaction with a sizable coupling may be constrained by the muonium-antimuonium oscillation~\cite{Pontecorvo:1957cp,Feinberg:1961zza,Lee:1977tib,Lee:1977qz,Willmann:1998gd} although we do not need a large coupling for our purpose.}. 
Interestingly, the parameter region predicts a large $d_{\mu}$ due to a small value of $m_{\varphi_1^+}$ and may be also explored by Higgs coupling measurements at future collider experiments. 
Ref.~\cite{deBlas:2019rxi} summarizes future sensitivities for the measurements of the SM Higgs couplings. 
In particular, the Future Circular Collider (FCC) may be able to measure $\kappa_{\mu}$ with relative precision of $\sim 0.4\%$ whose contours are shown as dot-dashed magenta lines in Fig.~\ref{fig:resMFmodel}~\footnote{Here, we just assume that the central value of $\kappa_{\mu}$ is $1$ at future collider experiments.}. 

In the whole parameter region shown in the figure, the muon EDM is predicted to be larger than the future sensitivity of the PSI muEDM experiment, $6 \times 10^{-23} \, e \, {\rm cm}$~\cite{Adelmann:2021udj,Sakurai:2022tbk,muonEDMinitiative:2022fmk}. 
The $2\sigma$ discrepancy of $(g-2)_{\mu}$ can be explained for $560 \, {\rm GeV} < m_{\psi_1} < 780 \, {\rm GeV}$, while only the region of $m_{\psi_1} \simeq m_{\varphi_1^+}$ is favored for the correct DM relic density. 
With the current parameter choice, the coannihilation process plays an important role in obtaining the correct relic density. 

For $m_{\psi_1} \simeq 850 \mathchar`- 860$ GeV, the DM sector contribution to the muon EDM accidentally disappears. 
This behavior can be understood as follows. 
In this region, $m_{LL} \simeq 1000$ GeV which means $m_{LL} \simeq m_{RR}$ for our current setup. 
Eq.~\eqref{eq:tantau} tells us that $\theta_{\rm phys} = 0$ or $| m_{LL} | = | m_{RR} |$ can lead to $\tan \tau = 0$, which makes $y_L^{i a}$ and $y_R^{i a}$ real. 
Therefore, there is no contribution to the muon EDM for $| m_{LL} | = | m_{RR} |$, even when the physical phase has a non-zero value, $\theta_{\rm phys} \neq 0$. 

We now comment on constraints from DM searches at colliders and DM direct and indirect detection experiments. 
\begin{itemize}
\item[1.] \underline{\bf Collider searches}\\
At the Large Hadron Collider (LHC), we expect a pair production of exotic charged scalars $\varphi_1$ decaying into muons and DM fermions:
\begin{align}
pp \to \varphi_1 \varphi_1 \to \mu \mu + \psi_1 \psi_1 \, ,
\end{align}
whose signal is two muons plus a large missing energy. 
The signal is similar to that of a pair production of sleptons decaying into leptons and a missing energy. 
Then, the ATLAS~\cite{ATLAS:2019lff,ATLAS:2019lng} and CMS~\cite{CMS:2018eqb} experiments put a lower bound on the DM mass. 
However, it is less than 500 GeV~\cite{ATLAS:2019lff}, which is outside the plot range of Fig.~\ref{fig:resMFmodel}. 
Ref.~\cite{Kawamura:2020qxo} has performed the numerical analysis to obtain a bound on the DM mass for the similar model, and it was found to be 200-300 GeV, depending on the size of the mixing angle $s_{\theta}$ in Eq.~\eqref{eq:Usdef} and the mass of $\varphi_1^+$. 
Ref.~\cite{ATLAS:2019lng} has investigated the case with $m_{\psi_1} \sim m_{\varphi_1^+}$ and put a lower bound on $m_{\varphi_1^+}$ less than 250 GeV for $m_{\varphi_1^+} - m_{\psi_1} = \mathcal{O}(10)$ GeV. 

\item[2.] \underline{\bf Indirect detection} \\
As mentioned in Sec.~\ref{DMpheno}, the annihilation cross section of our DM $\psi_1$ is dominated by $\psi_1 \psi_1 \to \mu \bar{\mu}$. 
For the parameter region in Fig.~\ref{fig:resMFmodel}, we obtain the prediction of $\langle \sigma v \rangle_{\mu \bar{\mu}} \simeq \mathcal{O}(10^{-27})$-$\mathcal{O}(10^{-28}) \, {\rm cm}^3/{\rm s}$. 
Ref.~\cite{Garny:2013ama} has studied a constraint on the annihilation cross section of a Majorana DM, whose annihilation modes are $\psi_1 \psi_1 \to \ell \bar{\ell} \gamma$ and $\psi_1 \psi_1 \to \gamma \gamma$. 
The combination of the thermally averaged cross sections, $\langle \sigma v \rangle_{\mu \bar{\mu} \gamma} + 2 \langle \sigma v \rangle_{\gamma \gamma}$, is constrained to be less than $10^{-26}$-$10^{-27} \, {\rm cm}^3/{\rm s}$, depending on the DM mass. 
The cross sections of these annihilation processes, however, are several orders of magnitude smaller than that of $\langle \sigma v \rangle_{\mu \bar{\mu}}$ in our model, and therefore, ref.~\cite{Garny:2013ama} does not put a constraint on the parameter region shown in Fig.~\ref{fig:resMFmodel}. 
It is notable that due to the gauge invariance, we should consider the $\psi_1 \psi_1 \to \ell \bar{\ell} \gamma$ process together with the $\psi_1 \psi_1 \to \ell \bar{\ell} Z$ process. 
Such processes may be able to be explored by the PAMELA anti-proton search~\cite{PAMELA:2010kea}, and there are studies for Majorana DM models~\cite{Garny:2011cj,Garny:2011ii}, although they indicate that it is difficult to observe a Majorana DM at current and future telescopes. 

\item[3.] \underline{\bf Direct detection}\\
Our Majorana DM scattering with the nucleon is induced by interactions presented in Eq.~\eqref{eq:Leff-DD}. 
For the current parameter set, we obtain the SI DM-nucleon scattering cross section of $\mathcal{O}(10^{-47}\mathchar`-10^{-50}) \, {\rm cm}^2$. 
This is smaller than the current limit from the LZ experiment~\cite{LZ:2022ufs}, which is $(1.5\mathchar`-2.4) \times 10^{-46} \, {\rm cm}^2$ for the DM mass range in Fig.~\ref{fig:resMFmodel}. 
The LZ experiment~\cite{LZ:2022ufs} also put constraints on the SD DM-proton and DM-neutron scattering cross sections, but both are weaker than that of the SI cross section, and therefore, no region of the parameter space is excluded by direct detection experiments. 
With the future sensitivity of the LZ experiment, the upper limit on the SI cross section will be improved by one order of magnitude~\cite{Mount:2017qzi,LZ:2019sgr}, which is still not sufficient to explore our parameter space. 
By the future sensitivity of PandaX-4T with 5.6 tonne$\cdot$year exposure~\cite{PandaX:2018wtu}, we may be able to explore the parameter space in Fig.~\ref{fig:resMFmodel}. 
Their current limit on the SI DM-nucleon scattering cross section~\cite{PandaX-4T:2021bab} can be read as $(2.6\mathchar`-4.2) \times 10^{-46} \, {\rm cm}^2$ for the DM mass range of $550 \, {\rm GeV} \leq m_{\psi_1} \leq 860 \, {\rm GeV}$, and hence, if the future limit is improved by a few orders of magnitude, a heavy DM mass region with $m_{\psi_1} \simeq m_{\varphi_1^+}$ will be explored at the PandaX-4T experiment. 

\end{itemize}

Finally, let us discuss how our new physics contributions to $\Delta a_{\mu}$ and $d_{\mu}$ depend on input parameter choices. 
First of all, a different choice of $\theta_{\rm phys}$ can change our predictions for $\Delta a_{\mu}$ and $d_{\mu}$ shown in Fig.~\ref{fig:resMFmodel}. 
It is expected that $d_{\mu}$ is maximized by choosing $\theta_{\rm phys} \sim \pi / 4$, because $\theta_{\rm phys} \to 0$ or $\theta_{\rm phys} \to \pi / 2$ leads to $d_{\mu} \approx 0$. 
On the other hand, the contribution to $\Delta a_{\mu}$ does not have such a clear dependence on $\theta_{\rm phys}$. 
Actually, both observables strongly depend on the input parameter set of $(m_D, m_{LL}, m_{RR}, \theta_{\rm phys})$. 
For example, if $(m_D, m_{LL}, m_{RR}) = (700 \, {\rm GeV}, 200 \, {\rm GeV}, 1000 \, {\rm GeV})$ are chosen, $\Delta a_{\mu} < 0$ for $0 < \theta_{\rm phys} \lesssim 0.42$ and the $2\sigma$ deviation can be explained for $\pi / 6 < \theta_{\rm phys} < \pi / 4$, and $d_{\mu}$ is maximized around $\theta_{\rm phys} \simeq \pi / 6$. 
Instead, if we choose $(m_D, m_{LL}, m_{RR}) = (500 \, {\rm GeV}, 990 \, {\rm GeV}, 1000 \, {\rm GeV})$, $\Delta a_{\mu} < 0$ is predicted in almost all range of $\theta_{\rm phys}$ and the $2\sigma$ deviation can be explained only around $\theta_{\rm phys} \simeq 1.45$, and the peak of $d_{\mu}$ appears at $\theta_{\rm phys} \simeq 1.35$. 
In any case, our prediction of the muon EDM is $d_{\mu} > \mathcal{O}(10^{-22})\, e \, {\rm cm}$. 
These observables also depend on the values of $M_{\phi}^2, M_{\eta}^2$. 
As one can see in Fig.~\ref{fig:resMFmodel}, the predictions of $\Delta a_{\mu}$ and $d_{\mu}$ become small as $m_{\varphi_1^+}$ increases. 
In contrast, a large $a$ enhances the contributions by a few \%. 
This is because $a$ is related to the difference between $m_{\varphi_1^+}^2$ and $m_{\varphi_2^+}^2$ (see Eqs.~\eqref{eq:eigenpm1} and \eqref{eq:eigenpm2}), and hence, a larger $a$ leads to a slightly smaller $m_{\varphi_1^+}^2$.

%#######################
\section{Conclusion}
\label{conclusion}

In the present paper, we have investigated a prediction for the muon EDM obtained in a model of DM. 
As shown in Table~\ref{tab:muon EDM prospects}, the radiative stability approach has a clear advantage to enhance the muon EDM, and we focused on a model in which the muon mass is generated radiatively. 
With appropriate discrete symmetries, exotic particles, $\psi$, $\phi$ and $\eta$, have couplings to the muon (and also to the SM Higgs doublet). 
In this model, one of the complex phases in the couplings cannot be removed by any field redefinition and provides a physical CP phase, which leads to a new contribution to the muon EDM. 
$\psi_{L, R}$ are singlet under the SM gauge groups and the lightest mode gives a candidate of the Majorana fermion DM. 

We found that even when the DM mass is heavier than the current collider bound, $m_{\psi_1} > 500$ GeV, the model predicts a muon EDM larger than $10^{-22} \, e \, {\rm cm}$ which can be tested at the PSI muEDM experiment. 
In the parameter space where the discrepancy of the muon $g-2$ and the correct DM relic density are explained at the same time, the model predicts $d_{\mu} \simeq (4 \mathchar`- 5) \times 10^{-22} \, e \, {\rm cm}$. 
For the case of $m_{\psi_1} > m_{\varphi_1^+}$, the muon EDM can be even larger, $d_{\mu} \simeq (7 \mathchar`- 8) \times 10^{-22} \, e \, {\rm cm}$, due to a small value of $m_{\varphi_1^+}$, although $\psi_1$ does not give a DM candidate. 
Furthermore, once we forget a new physics explanation for the muon $g-2$ discrepancy as well as the DM relic density, the muon EDM can be larger than the future sensitivities of the ongoing Fermilab Muon $g-2$ and projected J-PARC Muon $g-2$/EDM experiments. 

One of the most promising approaches to probe our DM model is a future muon collider (see e.g. ref.~\cite{AlAli:2021let} and references therein) because a muon collider is expected to have a new particle mass reach higher than that of the LHC and also our DM fermion directly couples to the muon. 
It would be interesting to explore the phenomenology of our DM model to generate the radiative muon mass, the muon $g-2$ and the muon EDM at a muon collider, which is left for future study.

%#######################
\section*{Acknowledgements}

KSK is supported by Natural Science Foundation of China (NSFC) under grant No. 12050410233. 
YN is supported by NSFC under grant No. 12150610465.

\appendix

%#######################
\section{Neutrino sector}
\label{app:Neutrino}

\subsection{A scalar triplet extension}
\label{app:Neutrino-triplet}

In our model, due to the muon number symmetry, we need a further extension for obtaining the correct neutrino mixing angles. 
One of the simplest way to reproduce them is to introduce a scalar triplet, as discussed in appendix~A of ref.~\cite{Abe:2017jqo}. 
At first, we can write down dimension-five operators which are related to lepton doublets as
\begin{align}
- \frac{c_{H H}^{a b}}{M_{H H}} \left( \overline{L_L^a} \widetilde{H} \right) \left( \widetilde{H}^T (L_L^c)^b \right) + {\rm h.c.} \, ,
\end{align}
where $\widetilde{H} = i \sigma_2 H^*$, and $a, b = e, \mu, \tau$ are indices for lepton species. 
Note that the exotic number symmetry forbids terms of $\left( \overline{L_L^a} \widetilde{H} \right) \left( \widetilde{\phi}^T (L_L^c)^b \right)$ and $\left( \overline{L_L^a} \widetilde{\phi} \right) \left( \widetilde{H}^T (L_L^c)^b \right)$, and the term of $\left( \overline{L_L^a} \widetilde{\phi} \right) \left( \widetilde{\phi}^T (L_L^c)^b \right)$ is irrelevant to the discussion on the neutrino mixing angles, because $\phi$ does not acquire a nonzero VEV. 
The coefficient $c_{H H}$ is a $3 \times 3$ matrix, but symmetries of the model makes it have the form,
\begin{align}
c_{H H} = \begin{pmatrix}
c_{H H}^{e e} & 0 & c_{H H}^{e \tau} \\
0 & c_{H H}^{\mu \mu} & 0 \\
c_{H H}^{\tau e} & 0 & c_{H H}^{\tau \tau}
\end{pmatrix} \, .
\end{align}
Therefore, the Pontecorvo-Maki-Nakagawa-Sakata (PMNS) matrix obtained from our model cannot be consistent with the experimental result at this stage. 
However, once we introduce a $SU(2)_L$ triplet scalar $\Delta$ to the model, which has the $U(1)_Y$ charge of $-1$ and odd muon number, we can write additional terms of
\begin{align}
- c_{\Delta}^{a b} \overline{L_L^a} \Delta (L_L^c)^b \, , \quad c_{\Delta} = \begin{pmatrix}
0 & c_{\Delta}^{e \mu} & 0 \\
c_{\Delta}^{\mu e} & 0 & c_{\Delta}^{\mu \tau} \\
0 & c_{\Delta}^{\tau \mu} & 0 \\
\end{pmatrix} \, .
\end{align}
When $\Delta$ acquires a nonzero VEV, $v_{\Delta} \neq 0$, we can reproduce all elements for the neutrino mass matrix as
\begin{align}
m_{\nu} = \begin{pmatrix}
(m_{\nu})^{e e} & (m_{\nu})^{e \mu} & (m_{\nu})^{e \tau} \\
(m_{\nu})^{\mu e} & (m_{\nu})^{\mu \mu} & (m_{\nu})^{\mu \tau} \\
(m_{\nu})^{\tau e} & (m_{\nu})^{\tau \mu} & (m_{\nu})^{\tau \tau}
\end{pmatrix} \, .
\label{eq:NeutrinoDelta}
\end{align}
Here, each $(m_{\nu})^{a b}$ is estimated by
\begin{align}
(m_{\nu})^{a b} = \begin{cases}
{\displaystyle \frac{c_{H H}^{a b}}{2} \frac{v_H^2}{M_{H H}}} & \text{for $(a, b) = (e, e), (e, \tau), (\mu, \mu), (\tau, e), (\tau, \tau)$} \\[2.4ex]
{\displaystyle c_{\Delta}^{a b} v_{\Delta}} & \text{for $(a, b) = (e, \mu), (\mu, e), (\mu, \tau), (\tau, \mu)$}
\end{cases} \, .
\end{align}
Then, if $v_{\Delta}$ is $\mathcal{O}(v_H^2 / M_{H H})$ with $c_{H H}^{a b} \sim c_{\Delta}^{a b}$, all elements of Eq.~\eqref{eq:NeutrinoDelta} have the similar order, and hence, the large mixing angles for the PMNS matrix can be obtained. 
For the neutrino masses of $\mathcal{O}({\rm eV})$, the mass scale for $M_{H H}$ is required to be $\mathcal{O}(10^{13 {\rm -} 14})$ GeV.

\subsection{Right-handed neutrinos}
\label{app:Neutrino-RHN}

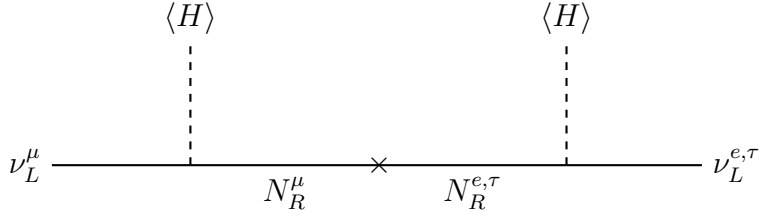
\begin{figure}[t]
\begin{center}
\begin{tikzpicture}
\begin{feynman}[large]
\vertex (a1) {\(\nu_L^{\mu}\)};
\vertex [right=2.2cm of a1] (b1);
\vertex [right=2.5cm of b1] (c1);
\vertex [right=2.5cm of c1] (d1);
\vertex [right=1.8cm of d1] (e1) {\(\nu_L^{e, \tau}\)};
\vertex [above=1.6cm of b1] (f1) {\(\langle H \rangle\)};
\vertex [above=1.6cm of d1] (g1) {\(\langle H \rangle\)};
\vertex [right=2.2cm of b1] (x1) {\(\times\)};
\diagram [medium] {
(a1) -- (b1) -- [edge label'=\(N_R^{\mu}\)] (c1) -- [edge label'=\(N_R^{e, \tau}\)] (d1) -- (e1),
(b1) -- [scalar] (f1),
(d1) -- [scalar] (g1),
};
\end{feynman}
\end{tikzpicture}
\end{center}
\vspace{-0.5cm}
\caption{A Feynman diagram for the neutrino mixing between $\nu_L^{\mu}$ and $\nu_L^{e, \tau}$ where
``$\times$" indicates the soft breaking mixing between $N_R^{\mu}$ and $N_R^{e, \tau}$.}
\label{fig:NeutrinoMixing}
\end{figure}

Another possibility to reproduce the correct PMNS matrix is to introduce three generations of the right-handed neutrinos (RHNs), denoted as $N_R^{e, \mu, \tau}$. 
Similar to the charged lepton sector, only $N_R^{\mu}$ is odd under the muon number $L_{\mu}$, and we have additional Dirac Yukawa couplings and Majorana mass terms for neutrinos as ($\ell^{(')} = e, \tau$)
\begin{align}
\mathcal{L}_N = - y_{\nu}^{\ell \ell'} \overline{L_L^{\ell}} \widetilde{H} N_R^{\ell'} - y_{\nu}^{\mu \mu} \overline{L_L^{\mu}} \widetilde{H} N_R^{\mu} - m_N^{\ell \ell'} \overline{N_R^{\ell \, c}} N_R^{\ell'} - m_N^{\mu \mu} \overline{N_R^{\mu \, c}} N_R^{\mu} - m_N^{\ell \mu} \overline{N_R^{\ell \, c}} N_R^{\mu} + {\rm h.c.} \, ,
\label{eq:Lneut}
\end{align}
where the last term breaks the $L_{\mu}$ symmetry softly, which is required for the correct neutrino mixing angles. 
This can be understood diagrammatically, as shown in Fig.~\ref{fig:NeutrinoMixing}. 
In addition to this diagram, the mixing between $\nu_L^e$ and $\nu_L^{\tau}$ is also induced
by the same diagram with changing $\mu \to e$ or $\tau$. 
The mass matrix of $\nu_L^{e, \mu, \tau}$ can be obtained as
\begin{align}
m_{\nu_L} = - \frac{v_H^2}{2} \begin{pmatrix}
y_{\nu}^{e e} & 0 & y_{\nu}^{e \tau} \\
0 & y_{\nu}^{\mu \mu} & 0 \\
y_{\nu}^{\tau e} & 0 & y_{\nu}^{\tau \tau} \\
\end{pmatrix}
\begin{pmatrix}
m_N^{e e} & m_N^{e \mu} & m_N^{e \tau} \\
m_N^{\mu e} & m_N^{\mu \mu} & m_N^{\mu \tau} \\
m_N^{\tau e} & m_N^{\tau \mu} & m_N^{\tau \tau} \\
\end{pmatrix}^{-1}
\begin{pmatrix}
y_{\nu}^{e e} & 0 & y_{\nu}^{\tau e} \\
0 & y_{\nu}^{\mu \mu} & 0 \\
y_{\nu}^{e \tau} & 0 & y_{\nu}^{\tau \tau} \\
\end{pmatrix} \, .
\end{align}
Therefore, if the mass scales of $m_N^{\ell \ell'}$ are similar with each other, we can obtain a full $3 \times 3$ matrix for light neutrino states, which can be consistent with experimental results on the PMNS matrix. 
Similar to the previous method, the mass scale of $m_N^{\ell \ell'}$ is required to be $\mathcal{O}(10^{13 {\rm -} 14})$ GeV for the neutrino masses of $\mathcal{O}({\rm eV})$ if $y_{\nu}^{\ell \ell'} \sim \mathcal{O}(1)$. 
%%%%%%%%%%%%%%%%%%%%%%%%%%%%%%%%%
%%%%%%%%%%%%%%%%%%%%%%%%%%%%%%%%%

\subsection{Comment on LFV}
\label{app:Neutrino-LFV}

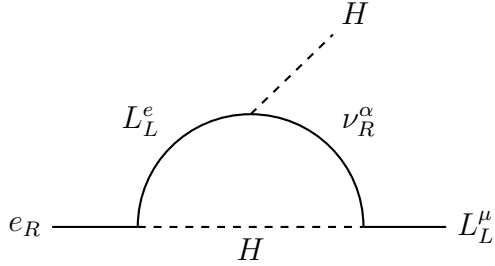
\begin{figure}[t]
\begin{center}
\begin{tikzpicture}
\begin{feynman}[large]
\vertex (a1) {\(e_R\)};
\vertex [right=1.5cm of a1] (b1);
\vertex [right=3cm of b1] (c1);
\vertex [right=1.1cm of c1] (d1) {\(L_L^{\mu}\)};
\vertex [above right=2.12132cm of b1] (f1);
\vertex [above right=1.5cm of f1] (g1) {\(H\)};
\diagram [medium] {
(a1) -- (b1) -- [scalar, edge label'=\(H\)] (c1) -- (d1),
(b1) -- [quarter left, edge label=\(L_L^e\)] (f1) -- [quarter left, edge label=\(\nu_R^{\alpha}\)] (c1),
(f1) -- [scalar] (g1),
};
\end{feynman}
\end{tikzpicture}
\end{center}
\vspace{-0.5cm}
\caption{One-loop contributions to off-diagonal elements of Yukawa couplings of charged leptons. 
Here, $\nu_R^{\alpha}$ is the mass eigenstate of RHNs.}
\label{fig:oneloopyell}
\end{figure}

For the second example, however, we have LFV processes due to the soft $L_{\mu}$ breaking terms. 
To see this, we focus on the first two generations of leptons, namely, the electron-muon system. 
From Eq.~\eqref{eq:Lneut} and the SM Yukawa interactions for the electron, we have one-loop contributions to off-diagonal elements of Yukawa couplings of charged leptons, as shown in Fig.~\ref{fig:oneloopyell}. 
%%%%%%%%%%%%%%%%%%%%%%%%%%%%%%%%%
%%%%%%%%%%%%%%%%%%%%%%%%%%%%%%%%%
Due to the soft $L_{\mu}$ breaking, the mass eigenstate $\nu_R^{\alpha}$ can be written by one mixing angle $\theta_N$ as
\begin{align}
\begin{pmatrix}
\nu_R^1 \\
\nu_R^2
\end{pmatrix} = \begin{pmatrix}
\cos \theta_N & \sin \theta_N \\
-\sin \theta_N & \cos \theta_N
\end{pmatrix} \begin{pmatrix}
N_R^e \\
N_R^{\mu}
\end{pmatrix} \, .
\end{align}
After integrating out the RHNs, we obtain the off-diagonal element,
\begin{align}
\mathcal{L}_{\rm eff} \supset - \delta y_{\mu e} \overline{L_L^{\mu}} H e_R + {\rm h.c.} \, ,
\end{align}
where $\delta y_{\mu e}$ is obtained by Fig.~\ref{fig:oneloopyell} and roughly estimated as
\begin{align}
\delta y_{\mu e} \simeq \frac{y_e}{16 \pi^2} y_{\nu}^{e e} y_{\nu}^{\mu \mu} \sin \theta_N \cos \theta_N \ln \frac{M_{\nu_R^1}^2}{M_{\nu_R^2}^2} \, ,
\end{align}
with $y_e$ being the SM Yukawa coupling of the term $\overline{L_L^e} H e_R$. 
This off-diagonal element can be removed by field redefinition of left-handed lepton doublets,
\begin{align}
\ell_L^e = L_L^e \cos \theta_{e \mu} + L_L^{\mu} \sin \theta_{e \mu} \, , ~ \ell_L^{\mu} = - L_L^e \sin \theta_{e \mu} + L_L^{\mu} \cos \theta_{e \mu} \, ,
\label{eq:emuredef}
\end{align}
which lead to an electron coupling with exotic particles,
\begin{align}
y_{\phi} \overline{L_L^{\mu}} \phi^{\dagger} \psi_R \to y_{\phi}^e \overline{\ell_L^e} \phi^{\dagger} \psi_R + y_{\phi}^{\mu} \overline{\ell_L^{\mu}} \phi^{\dagger} \psi_R \, ,
\end{align}
where we define $y_{\phi}^e \equiv y_{\phi} \sin \theta_{e \mu}$ and $y_{\phi}^{\mu} \equiv y_{\phi} \cos \theta_{e \mu}$. 
Then, we have a one-loop contribution to $\mu \to e \gamma$ by replacing $\mu_L \to e_L$ in the diagram (b) of Fig.~\ref{fig:MFdiagrams}. 
Its branching ratio can be calculated as~\cite{Lindner:2016bgg}
\begin{align}
{\rm BR}(\mu \to e \gamma) = \frac{3 (4 \pi)^3 \alpha}{2 G_F^2 m_{\mu}^2} \Bigl( \left| C_T^{e \mu} \right|^2 + \left| C_{T'}^{e \mu} \right|^2 \Bigr) \, {\rm BR}(\mu \to e \nu_{\mu} \bar{\nu}_e) \, ,
\label{eq:BRmuegamma}
\end{align}
where $G_F$ is the Fermi constant, BR$(\mu \to e \nu_{\mu} \bar{\nu}_e) \approx 1$, and $C_T^{e \mu}$ and $C_{T'}^{e \mu}$ are coefficients of dipole operators, whose definitions are
\begin{align}
\mathcal{L}_{\rm LFV} \supset - \frac{e}{2} C_T^{e \mu} \left( \bar{e} \sigma^{\alpha \beta} \mu \right) F_{\alpha \beta} - \frac{e}{2} C_{T'}^{e \mu} \left( \bar{e} i \sigma^{\alpha \beta} \gamma_5 \mu \right) F_{\alpha \beta} \, .
\label{eq:LFVOPs}
\end{align}
From Eqs.~\eqref{eq:chiralrotmu}, \eqref{eq:CT0} and \eqref{eq:CTp0}, leading contributions to $C_T^{e \mu}$ and $C_{T'}^{e \mu}$ can be easily estimated by replacing $y_L^{i a} \to y_L^{i a} \sin \theta_{e \mu}$. 
Similar to this replacement, our predictions of $a_{\mu}$ and $d_{\mu}$ in Eqs.~\eqref{eq:amu} and \eqref{eq:dmu} are changed as
\begin{align}
a_{\mu} &= 2 m_{\mu} \left( C_T (0) \cos \theta_{\mu} + C_{T'} (0) \sin \theta_{\mu} \right) \cos \theta_{e \mu} \, , \label{eq:amu2} \\
d_{\mu} &= e \left( C_{T'} (0) \cos \theta_{\mu} - C_T (0) \sin \theta_{\mu} \right) \cos \theta_{e \mu} \, . \label{eq:dmu2}
\end{align}
From these facts, we have the following relation between $C_T^{e \mu}$, $C_{T'}^{e \mu}$ and $a_{\mu}$, $d_{\mu}$ as
\begin{align}
|C_T^{e \mu}|^2 + |C_{T'}^{e \mu}|^2 &= \left( \left| C_T(0) c_{\theta_{\mu}/2} + C_{T'}(0) s_{\theta_{\mu}/2} \right|^2 + \left| C_{T'}(0) c_{\theta_{\mu}/2} - C_T(0) s_{\theta_{\mu}/2} \right|^2 \right) \sin^2 \theta_{e \mu} \nonumber \\[0.3ex]
&= \left( |C_T(0)|^2 + |C_{T'}(0)|^2 \right) \sin^2 \theta_{e \mu} \nonumber \\[0.3ex]
&= \left( \left| \frac{a_{\mu}}{2 m_{\mu}} \right|^2 + \left| \frac{d_{\mu}}{e} \right|^2 \right) \tan^2 \theta_{e \mu} \, ,
\end{align}
and hence, the branching ratio in Eq.~\eqref{eq:BRmuegamma} can be expressed by $a_{\mu}$ and $d_{\mu}$ as
\begin{align}
{\rm BR}(\mu \to e \gamma) = \frac{3 (4 \pi)^3 \alpha}{2 G_F^2 m_{\mu}^2} \left( \left| \frac{a_{\mu}}{2 m_{\mu}} \right|^2 + \left| \frac{d_{\mu}}{e} \right|^2 \right) \tan^2 \theta_{e \mu} \, .
\label{eq:BRmuegamma2}
\end{align}
By assuming $a_{\mu} = 2.51 \times 10^{-9}$ and $d_{\mu} = 4.5 \times 10^{-22} \, e \, {\rm cm}$ as we found in our model, the upper limit on $\tan \theta_{e \mu}$ can be obtained as
\begin{align}
\tan \theta_{e \mu} \lesssim 6.7 \times 10^{-6} \, ,
\label{eq:tanthetalimit}
\end{align}
where we have used the current upper bound on the branching ratio, BR$(\mu \to e \gamma) < 4.2 \times 10^{-13}$~\cite{MEG:2016leq}. 
Then, the mixing angle in Eq.~\eqref{eq:emuredef} should be tiny, which means a small off-diagonal element compared to $y_e$, $\delta y_{\mu e} \ll y_e$. 
To realize the constraint in Eq.~\eqref{eq:tanthetalimit}, we roughly need $\delta y_{\mu e} / y_e \sim 6 \times 10^{-6}$. 
Assuming $\ln (M_{\nu_R^1}^2 / M_{\nu_R^2}^2) \sim \mathcal{O}(1)$ and $\sin \theta_N \sim \cos \theta_N \sim 1/\sqrt{2}$ due to the similar order for all $m_N^{\ell \ell'}$ to obtain large mixing angles for neutrinos, we obtain $y_{\nu}^{e e} y_{\nu}^{\mu \mu} \sim 2 \times 10^{-3}$, and the scale of RHNs will be $\mathcal{O}(10^{10})$ GeV for the light neutrino masses of $\mathcal{O}({\rm eV})$.

%#######################
\section{Loop integrals}
\label{app:LoopIntegrals}

In the appendix, we summarize loop integrals relevant to our analysis. 
Note that we use dimensional regularization, and $\Delta_{\epsilon} \equiv \frac{2}{\epsilon} - \gamma_E + \ln 4\pi$ with the Euler constant $\gamma_E$ and $\epsilon = 4 - D$ diverges when $D \to 4$. 

\begin{itemize}
\item[] -- Self-energy integral --
\begin{align}
B_0 (p^2, m_0^2, m_1^2) &= \Delta_{\epsilon} - \int_0^1 \! dx \ln \left[ \frac{- x (1 - x) p^2 + x m_1^2 + (1 - x) m_0^2}{\mu^2} \right] \, .
\end{align}
When $p^2 = 0$ and $m_0^2 = m_1^2 \equiv m^2$, we can obtain the following simple form:
\begin{align}
B_0 (0, m^2, m^2) &= \Delta_{\epsilon} - \ln \frac{m^2}{\mu^2} \, .
\end{align}
Note that $B_0$ has a divergent part which should be cancelled for physical predictions. 

\item[] -- Triangle integrals --
\begin{align}
C_0 (p_1^2, p_{12}^2, p_2^2, m_0^2, m_1^2, m_2^2) &= - \int_0^1 \! dx_1 \int_0^{1 - x_1} \! dx_2 \frac{1}{C_3} \, , \\[0.5ex]
C_1 (p_1^2, p_{12}^2, p_2^2, m_0^2, m_1^2, m_2^2) &= \int_0^1 \! dx_1 \int_0^{1 - x_1} \! dx_2 \frac{x_1}{C_3} \, , \\[0.5ex]
C_{ij} (p_1^2, p_{12}^2, p_2^2, m_0^2, m_1^2, m_2^2) &= - \int_0^1 \! dx_1 \int_0^{1 - x_1} \! dx_2 \frac{x_i x_j}{C_3} \, ,
\end{align}
where $C_3 = x_1 (x_1 - 1) p_1^2 + x_2 (x_2 - 1) p_2^2 + x_1 x_2 (p_1^2 + p_2^2 - p_{12}^2) + x_1 m_1^2 + x_2 m_2^2 + (1 - x_1 - x_2) m_0^2$ with $p_{12}^2 = (p_1 - p_2)^2$. 
When all of $p_1^2$, $p_2^2$ and $p_{12}^2$ are $0$, these integrals are simplified as follows. 
First, the $C_0$ function is (here we use $r_i \equiv m_i^2 / m_0^2$):
\begin{align}
C_0 (0, 0, 0, m_0^2, m_1^2, m_2^2) &= \frac{1}{m_0^2} \frac{r_1 \ln r_1 - r_1 r_2 \ln r_1 - r_2 \ln r_2 + r_1 r_2 \ln r_2}{(1 - r_1) (1 - r_2) (r_1 - r_2)} \, , \\[0.5ex]
C_0 (0, 0, 0, m_0^2, m_1^2, m_1^2) &= \frac{1}{m_0^2} \frac{1 - r_1 + \ln r_1}{(1 - r_1)^2} \, , \\[0.5ex]
C_0 (0, 0, 0, m_0^2, m_0^2, m_0^2) &= - \frac{1}{2 m_0^2} \, .
\end{align}
Note that $C_0 (0, 0, 0, m_0^2, m_1^2, m_2^2)$ is symmetric under exchanging two of $m_i^2$, e.g., $C_0 (0, 0, 0, m_0^2, m_1^2, m_2^2) = C_0 (0, 0, 0, m_1^2, m_0^2, m_2^2) = C_0 (0, 0, 0, m_0^2, m_2^2, m_1^2)$. 

Second, the $C_1$ function is simplified as 
\begin{align}
C_1 (0, 0, 0, m_0^2, m_1^2, m_2^2) &= - \frac{1}{2 m_0^2} \left[ \frac{r_1}{(1 - r_1) (r_1 - r_2)} + \frac{r_1 (r_1 - 2 r_2 + r_1 r_2)}{(1 - r_1)^2 (r_1 - r_2)^2} \ln r_1 \right. \nonumber \\
&\hspace{6.0em} \left. + \frac{r_2^2}{(1 - r_2) (r_1 - r_2)^2} \ln r_2  \right] \, , \\[0.5ex]
C_1 (0, 0, 0, m_0^2, m_1^2, m_1^2) &= - \frac{1}{4 m_0^2} \frac{r_1^2 - 4 r_1 + 3 + 2 \ln r_1}{(1 - r_1)^3} \, , \\[0.5ex]
C_1 (0, 0, 0, m_0^2, m_0^2, m_0^2) &= \frac{1}{6 m_0^2} \, .
\end{align}
Note that there is a symmetric property only for $m_0^2 \leftrightarrow m_2^2$,
$C_1 (0, 0, 0, m_0^2, m_1^2, m_2^2) = C_1 (0, 0, 0, m_2^2, m_1^2, m_0^2)$. 

Third, the $C_{11}$ function is
\begin{align}
C_{11} (0, 0, 0, m_0^2, m_1^2, m_2^2) &= - \frac{1}{3 m_0^2} \left[ \frac{r_1 (r_1^2 - 3 r_1 + 5 r_2 - 3 r_1 r_2)}{2 (1 - r_1)^2 (r_1 - r_2)^2} \right. \nonumber \\
&\hspace{4.0em} - \frac{r_1 (r_1^2 r_2^2 + r_1^2 r_2 + r_1^2 - 3 r_1 r_2^2 - 3 r_1 r_2 + 3 r_2^2)}{(1 - r_1)^3 (r_1 - r_2)^3} \ln r_1 \nonumber \\
&\hspace{4.0em} \left. + \frac{r_2^3}{(1 - r_2) (r_1 - r_2)^3} \ln r_2 \right] \, , \\[0.5ex]
C_{11} (0, 0, 0, m_0^2, m_1^2, m_1^2) &= - \frac{1}{18 m_0^2} \frac{2 r_1^3 - 9 r_1^2 + 18 r_1 - 11 - 6 \ln r_1}{(1 - r_1)^4} \, , \\[0.5ex]
C_{11} (0, 0, 0, m_0^2, m_0^2, m_0^2) &= - \frac{1}{12 m_0^2} \, .
\end{align}

Finally, the $C_{12}$ function is
\begin{align}
C_{12} (0, 0, 0, m_0^2, m_1^2, m_2^2) &= - \frac{1}{6 m_0^2} \left[ \frac{r_1^2 r_2 - r_1^2 + r_1 r_2^2 - r_2^2}{(1 - r_1) (1 - r_2) (r_1 - r_2)^2} \right. \nonumber \\
&\hspace{6.0em} - \frac{r_1^2 (2 r_1 r_2 + r_1 - 3 r_2)}{(1 - r_1)^2 (r_1 - r_2)^3} \ln r_1 \nonumber \\
&\hspace{6.0em} \left. + \frac{r_2^2 (2 r_1 r_2 - 3 r_1 + r_2)}{(1 - r_2)^2 (r_1 - r_2)^3} \ln r_2  \right] \, , \\[0.5ex]
C_{12} (0, 0, 0, m_0^2, m_1^2, m_1^2) &= - \frac{1}{36 m_0^2} \frac{2 r_1^3 - 9 r_1^2 + 18 r_1 - 11 - 6 \ln r_1}{(1 - r_1)^4} \, , \\[0.5ex]
C_{12} (0, 0, 0, m_0^2, m_0^2, m_0^2) &= - \frac{1}{24 m_0^2} \, .
\end{align}

\end{itemize}

%#######################
\section{Full forms for $y_{\mu}^{\rm eff}$, $C_T$ and $C_{T'}$}
\label{app:FullForms}

We here present full expressions for $y_{\mu}^{\rm eff}$, $C_T$ and $C_{T'}$ at the one-loop order. 
In their calculations, \texttt{FeynCalc}~\cite{{Mertig:1990an,Shtabovenko:2016sxi,Shtabovenko:2020gxv}} is used, and to deal with $\gamma_5$ in dimensional regularization, we adopt the 't~Hooft-Veltman-Breitenlohner-Maison (BMHV) prescription~\cite{tHooft:1972tcz,Breitenlohner:1977hr}. 
The results are
\begin{align}
y_{\mu}^{\rm eff} (p_{h^0}^2) &= \sum_{i, j, a} \Biggl\{ \Biggl. - \frac{y_L^{i a} y_R^{j a} A_{i j}}{16 \pi^2} m_{\psi_a} C_0 (m_{\mu}^2, m_{\mu}^2, p_{h^0}^2, m_{\varphi_i^+}^2, m_{\psi_a}^2, m_{\varphi_j^+}^2) \nonumber \\
&\hspace{4.0em} + \frac{A_{ij}}{16 \pi^2} m_{\mu}^{\rm rad} \left[ y_R^{i a \, *} y_R^{j a} C_1 (m_{\mu}^2, p_{h^0}^2, m_{\mu}^2, m_{\psi_a}^2, m_{\varphi_i^+}^2, m_{\varphi_j^+}^2) \right. \nonumber \\
&\hspace{10.0em} \left. + y_L^{i a} y_L^{j a \, *} C_2 (m_{\mu}^2, p_{h^0}^2, m_{\mu}^2, m_{\psi_a}^2, m_{\varphi_i^+}^2, m_{\varphi_j^+}^2) \right] \Biggl. \Biggr\} \, , \label{eq:YeffFull} \\[0.5ex]
C_T (q^2) &= \sum_{i, a} \Biggl\{ \Biggr. \frac{{\rm Re}[y_L^{i a} y_R^{i a}]}{16 \pi^2} m_{\psi_a} \Bigl[ \Bigr. Q_S \left( C_0 (q^2, m_{\psi_a}^2, m_{\varphi_i^+}^2) + 2 C_1 (q^2, m_{\psi_a}^2, m_{\varphi_i^+}^2) \right) \nonumber \\[0.3ex]
&\hspace{14.0em} - 2 Y_{\psi} C_1 (q^2, m_{\varphi_i^+}^2, m_{\psi_a}^2) \Bigl. \Bigr] \nonumber \\[0.5ex]
&\hspace{4.0em} - \frac{\left| y_L^{i a} \right|^2 + \left| y_R^{i a} \right|^2}{16 \pi^2} {\rm Re}[m_{\mu}^{\rm rad}] \Bigl[ \Bigr. Q_S C_{\rm sub} (q^2, m_{\psi_a}^2, m_{\varphi_i^+}^2) \nonumber \\[0.3ex]
&\hspace{18.0em} + Y_{\psi} C_{\rm sub} (q^2, m_{\varphi_i^+}^2, m_{\psi_a}^2) \Bigl. \Bigr] \Biggl. \Biggr\} \, , \label{eq:CTfull} \\[1.0ex]
C_{T'} (q^2) &= \sum_{i, a} \Biggl\{ \Biggr. \frac{{\rm Im}[y_L^{i a} y_R^{i a}]}{16 \pi^2} m_{\psi_a} \Bigl[ \Bigr. Q_S \left( C_0 (q^2, m_{\psi_a}^2, m_{\varphi_i^+}^2) + 2 C_1 (q^2, m_{\psi_a}^2, m_{\varphi_i^+}^2) \right) \nonumber \\[0.3ex]
&\hspace{14.0em} - 2 Y_{\psi} C_1 (q^2, m_{\varphi_i^+}^2, m_{\psi_a}^2) \Bigl. \Bigr] \nonumber \\[0.5ex]
&\hspace{4.0em} - \frac{\left| y_L^{i a} \right|^2 + \left| y_R^{i a} \right|^2}{16 \pi^2} {\rm Im}[m_{\mu}^{\rm rad}] \Bigl[ \Bigr. Q_S C_{\rm sub} (q^2, m_{\psi_a}^2, m_{\varphi_i^+}^2) \nonumber \\[0.3ex]
&\hspace{18.0em} + Y_{\psi} C_{\rm sub} (q^2, m_{\varphi_i^+}^2, m_{\psi_a}^2) \Bigl. \Bigr] \Biggl. \Biggr\} \, , \label{eq:CTpfull}
\end{align}
where $Q_S = 1 + Y_{\psi}$, and we have defined
\begin{align}
C_0 (q^2, m_{\psi_a}^2, m_{\varphi_i^+}^2) &\equiv C_0 (m_{\mu}^2, m_{\mu}^2, q^2, m_{\varphi_i^+}^2, m_{\psi_a}^2, m_{\varphi_i^+}^2) \, , \\[0.5ex]
C_N (q^2, m_A^2, m_B^2) &\equiv C_N (m_{\mu}^2, q^2, m_{\mu}^2, m_A^2, m_B^2, m_B^2) \qquad (N = 1, 11, 12) \, ,
\end{align}
and for sub-leading contributions, we have also defined
\begin{align}
C_{\rm sub} (q^2, m_A^2, m_B^2) &\equiv C_1 (q^2, m_A^2, m_B^2) + C_{11} (q^2, m_A^2, m_B^2) + C_{12} (q^2, m_A^2, m_B^2) \, .
\end{align}

\bibliographystyle{utphys}
\bibliography{Radiative_muon}

\providecommand{\href}[2]{#2}\begingroup\raggedright\begin{thebibliography}{100}

\bibitem{Muong-2:2006rrc}
{\bfseries Muon g-2} Collaboration, G.~W. Bennett {\em et~al.}, ``{Final Report
  of the Muon E821 Anomalous Magnetic Moment Measurement at BNL},''
  \href{http://dx.doi.org/10.1103/PhysRevD.73.072003}{{\em Phys. Rev. D}
  {\bfseries 73} (2006) 072003},
  \href{http://arxiv.org/abs/hep-ex/0602035}{{\ttfamily arXiv:hep-ex/0602035}}.

\bibitem{Keshavarzi:2018mgv}
A.~Keshavarzi, D.~Nomura, and T.~Teubner, ``{Muon $g-2$ and $\alpha(M_Z^2)$: a
  new data-based analysis},''
  \href{http://dx.doi.org/10.1103/PhysRevD.97.114025}{{\em Phys. Rev. D}
  {\bfseries 97} no.~11, (2018) 114025},
  \href{http://arxiv.org/abs/1802.02995}{{\ttfamily arXiv:1802.02995
  [hep-ph]}}.

\bibitem{Aoyama:2020ynm}
T.~Aoyama {\em et~al.}, ``{The anomalous magnetic moment of the muon in the
  Standard Model},''
  \href{http://dx.doi.org/10.1016/j.physrep.2020.07.006}{{\em Phys. Rept.}
  {\bfseries 887} (2020) 1--166},
  \href{http://arxiv.org/abs/2006.04822}{{\ttfamily arXiv:2006.04822
  [hep-ph]}}.

\bibitem{Muong-2:2021ojo}
{\bfseries Muon g-2} Collaboration, B.~Abi {\em et~al.}, ``{Measurement of the
  Positive Muon Anomalous Magnetic Moment to 0.46 ppm},''
  \href{http://dx.doi.org/10.1103/PhysRevLett.126.141801}{{\em Phys. Rev.
  Lett.} {\bfseries 126} no.~14, (2021) 141801},
  \href{http://arxiv.org/abs/2104.03281}{{\ttfamily arXiv:2104.03281
  [hep-ex]}}.

\bibitem{Keshavarzi:2021eqa}
A.~Keshavarzi, K.~S. Khaw, and T.~Yoshioka, ``{Muon $g-2$: A review},''
  \href{http://dx.doi.org/10.1016/j.nuclphysb.2022.115675}{{\em Nucl. Phys. B}
  {\bfseries 975} (2022) 115675},
  \href{http://arxiv.org/abs/2106.06723}{{\ttfamily arXiv:2106.06723
  [hep-ex]}}.

\bibitem{Muong-2:2008ebm}
{\bfseries Muon (g-2)} Collaboration, G.~W. Bennett {\em et~al.}, ``{An
  Improved Limit on the Muon Electric Dipole Moment},''
  \href{http://dx.doi.org/10.1103/PhysRevD.80.052008}{{\em Phys. Rev. D}
  {\bfseries 80} (2009) 052008},
  \href{http://arxiv.org/abs/0811.1207}{{\ttfamily arXiv:0811.1207 [hep-ex]}}.

\bibitem{Ema:2021jds}
Y.~Ema, T.~Gao, and M.~Pospelov, ``{Improved Indirect Limits on Muon Electric
  Dipole Moment},''
  \href{http://dx.doi.org/10.1103/PhysRevLett.128.131803}{{\em Phys. Rev.
  Lett.} {\bfseries 128} no.~13, (2022) 131803},
  \href{http://arxiv.org/abs/2108.05398}{{\ttfamily arXiv:2108.05398
  [hep-ph]}}.

\bibitem{Chislett:2016jau}
{\bfseries Muon g-2} Collaboration, R.~Chislett, ``{The muon EDM in the g-2
  experiment at Fermilab},''
  \href{http://dx.doi.org/10.1051/epjconf/201611801005}{{\em EPJ Web Conf.}
  {\bfseries 118} (2016) 01005}.

\bibitem{Abe:2019thb}
M.~Abe {\em et~al.}, ``{A New Approach for Measuring the Muon Anomalous
  Magnetic Moment and Electric Dipole Moment},''
  \href{http://dx.doi.org/10.1093/ptep/ptz030}{{\em PTEP} {\bfseries 2019}
  no.~5, (2019) 053C02}, \href{http://arxiv.org/abs/1901.03047}{{\ttfamily
  arXiv:1901.03047 [physics.ins-det]}}.

\bibitem{Adelmann:2021udj}
A.~Adelmann {\em et~al.}, ``{Search for a muon EDM using the frozen-spin
  technique},'' \href{http://arxiv.org/abs/2102.08838}{{\ttfamily
  arXiv:2102.08838 [hep-ex]}}.

\bibitem{Sakurai:2022tbk}
M.~Sakurai {\em et~al.}, ``{muEDM: Towards a search for the muon electric
  dipole moment at PSI using the frozen-spin technique},'' in {\em {24th
  International Symposium on Spin Physics}}.
\newblock 1, 2022.
\newblock \href{http://arxiv.org/abs/2201.06561}{{\ttfamily arXiv:2201.06561
  [hep-ex]}}.

\bibitem{muonEDMinitiative:2022fmk}
{\bfseries muon EDM initiative} Collaboration, K.~S. Khaw {\em et~al.},
  ``{Search for the muon electric dipole moment using frozen-spin technique at
  PSI},'' \href{http://dx.doi.org/10.22323/1.402.0136}{{\em PoS} {\bfseries
  NuFact2021} (2022) 136}, \href{http://arxiv.org/abs/2201.08729}{{\ttfamily
  arXiv:2201.08729 [hep-ex]}}.

\bibitem{Cesarotti:2018huy}
C.~Cesarotti, Q.~Lu, Y.~Nakai, A.~Parikh, and M.~Reece, ``{Interpreting the
  Electron EDM Constraint},''
  \href{http://dx.doi.org/10.1007/JHEP05(2019)059}{{\em JHEP} {\bfseries 05}
  (2019) 059}, \href{http://arxiv.org/abs/1810.07736}{{\ttfamily
  arXiv:1810.07736 [hep-ph]}}.

\bibitem{Nakai:2016atk}
Y.~Nakai and M.~Reece, ``{Electric Dipole Moments in Natural Supersymmetry},''
  \href{http://dx.doi.org/10.1007/JHEP08(2017)031}{{\em JHEP} {\bfseries 08}
  (2017) 031}, \href{http://arxiv.org/abs/1612.08090}{{\ttfamily
  arXiv:1612.08090 [hep-ph]}}.

\bibitem{Ibrahim:1997gj}
T.~Ibrahim and P.~Nath, ``{The Neutron and the electron electric dipole moment
  in N=1 supergravity unification},''
  \href{http://dx.doi.org/10.1103/PhysRevD.58.019901}{{\em Phys. Rev. D}
  {\bfseries 57} (1998) 478--488},
  \href{http://arxiv.org/abs/hep-ph/9708456}{{\ttfamily arXiv:hep-ph/9708456}}.
  [Erratum: Phys.Rev.D 58, 019901 (1998), Erratum: Phys.Rev.D 60, 079903
  (1999), Erratum: Phys.Rev.D 60, 119901 (1999)].

\bibitem{Feng:2003mg}
T.-F. Feng, T.~Huang, X.-Q. Li, X.-M. Zhang, and S.-M. Zhao, ``{Lepton dipole
  moments and rare decays in the CP violating MSSM with nonuniversal soft
  supersymmetry breaking},''
  \href{http://dx.doi.org/10.1103/PhysRevD.68.016004}{{\em Phys. Rev. D}
  {\bfseries 68} (2003) 016004},
  \href{http://arxiv.org/abs/hep-ph/0305290}{{\ttfamily arXiv:hep-ph/0305290}}.

\bibitem{Feng:2006ei}
T.-F. Feng, X.-Q. Li, L.~Lin, J.~Maalampi, and H.-S. Song, ``{The Two-loop
  supersymmetric corrections to lepton anomalous magnetic and electric dipole
  moments},'' \href{http://dx.doi.org/10.1103/PhysRevD.73.116001}{{\em Phys.
  Rev. D} {\bfseries 73} (2006) 116001},
  \href{http://arxiv.org/abs/hep-ph/0604171}{{\ttfamily arXiv:hep-ph/0604171}}.

\bibitem{Feng:2008cn}
T.-F. Feng, L.~Sun, and X.-Y. Yang, ``{Electroweak and supersymmetric two-loop
  corrections to lepton anomalous magnetic and electric dipole moments},''
  \href{http://dx.doi.org/10.1016/j.nuclphysb.2008.03.019}{{\em Nucl. Phys. B}
  {\bfseries 800} (2008) 221--252},
  \href{http://arxiv.org/abs/0805.1122}{{\ttfamily arXiv:0805.1122 [hep-ph]}}.

\bibitem{Zhao:2014vga}
S.-M. Zhao, T.-F. Feng, X.-J. Zhan, H.-B. Zhang, and B.~Yan, ``{The study of
  lepton EDM in CP violating BLMSSM},''
  \href{http://dx.doi.org/10.1007/JHEP07(2015)124}{{\em JHEP} {\bfseries 07}
  (2015) 124}, \href{http://arxiv.org/abs/1411.4210}{{\ttfamily arXiv:1411.4210
  [hep-ph]}}.

\bibitem{Su:2022vju}
L.-H. Su, D.~He, X.-X. Dong, T.-F. Feng, and S.-M. Zhao, ``{Study of lepton
  EDMs in the U(1)$_{X}$ SSM *},''
  \href{http://dx.doi.org/10.1088/1674-1137/ac6e35}{{\em Chin. Phys. C}
  {\bfseries 46} no.~9, (2022) 093103},
  \href{http://arxiv.org/abs/2201.00517}{{\ttfamily arXiv:2201.00517
  [hep-ph]}}.

\bibitem{Nakai:2022vgp}
Y.~Nakai, R.~Sato, and Y.~Shigekami, ``{Muon electric dipole moment as a probe
  of flavor-diagonal CP violation},''
  \href{http://dx.doi.org/10.1016/j.physletb.2022.137194}{{\em Phys. Lett. B}
  {\bfseries 831} (2022) 137194},
  \href{http://arxiv.org/abs/2204.03183}{{\ttfamily arXiv:2204.03183
  [hep-ph]}}.

\bibitem{Hiller:2010ib}
G.~Hiller, K.~Huitu, T.~Ruppell, and J.~Laamanen, ``{A Large Muon Electric
  Dipole Moment from Flavor?},''
  \href{http://dx.doi.org/10.1103/PhysRevD.82.093015}{{\em Phys. Rev. D}
  {\bfseries 82} (2010) 093015},
  \href{http://arxiv.org/abs/1008.5091}{{\ttfamily arXiv:1008.5091 [hep-ph]}}.

\bibitem{Omura:2015xcg}
Y.~Omura, E.~Senaha, and K.~Tobe, ``{$\tau$- and $\mu$-physics in a general two
  Higgs doublet model with $\mu-\tau$ flavor violation},''
  \href{http://dx.doi.org/10.1103/PhysRevD.94.055019}{{\em Phys. Rev. D}
  {\bfseries 94} no.~5, (2016) 055019},
  \href{http://arxiv.org/abs/1511.08880}{{\ttfamily arXiv:1511.08880
  [hep-ph]}}.

\bibitem{Abe:2019bkf}
Y.~Abe, T.~Toma, and K.~Tsumura, ``{A $\mu$-$\tau$-philic scalar doublet under
  $Z_n$ flavor symmetry},''
  \href{http://dx.doi.org/10.1007/JHEP06(2019)142}{{\em JHEP} {\bfseries 06}
  (2019) 142}, \href{http://arxiv.org/abs/1904.10908}{{\ttfamily
  arXiv:1904.10908 [hep-ph]}}.

\bibitem{Hou:2021zqq}
W.-S. Hou, G.~Kumar, and S.~Teunissen, ``{Charged lepton EDM with extra Yukawa
  couplings},'' \href{http://dx.doi.org/10.1007/JHEP01(2022)092}{{\em JHEP}
  {\bfseries 01} (2022) 092}, \href{http://arxiv.org/abs/2109.08936}{{\ttfamily
  arXiv:2109.08936 [hep-ph]}}.

\bibitem{Cheung:2001ip}
K.-m. Cheung, ``{Muon anomalous magnetic moment and leptoquark solutions},''
  \href{http://dx.doi.org/10.1103/PhysRevD.64.033001}{{\em Phys. Rev. D}
  {\bfseries 64} (2001) 033001},
  \href{http://arxiv.org/abs/hep-ph/0102238}{{\ttfamily arXiv:hep-ph/0102238}}.

\bibitem{Arnold:2013cva}
J.~M. Arnold, B.~Fornal, and M.~B. Wise, ``{Phenomenology of scalar
  leptoquarks},'' \href{http://dx.doi.org/10.1103/PhysRevD.88.035009}{{\em
  Phys. Rev. D} {\bfseries 88} (2013) 035009},
  \href{http://arxiv.org/abs/1304.6119}{{\ttfamily arXiv:1304.6119 [hep-ph]}}.

\bibitem{Dorsner:2016wpm}
I.~Dor\v{s}ner, S.~Fajfer, A.~Greljo, J.~F. Kamenik, and N.~Ko\v{s}nik,
  ``{Physics of leptoquarks in precision experiments and at particle
  colliders},'' \href{http://dx.doi.org/10.1016/j.physrep.2016.06.001}{{\em
  Phys. Rept.} {\bfseries 641} (2016) 1--68},
  \href{http://arxiv.org/abs/1603.04993}{{\ttfamily arXiv:1603.04993
  [hep-ph]}}.

\bibitem{Dekens:2018bci}
W.~Dekens, J.~de~Vries, M.~Jung, and K.~K. Vos, ``{The phenomenology of
  electric dipole moments in models of scalar leptoquarks},''
  \href{http://dx.doi.org/10.1007/JHEP01(2019)069}{{\em JHEP} {\bfseries 01}
  (2019) 069}, \href{http://arxiv.org/abs/1809.09114}{{\ttfamily
  arXiv:1809.09114 [hep-ph]}}.

\bibitem{Altmannshofer:2020ywf}
W.~Altmannshofer, S.~Gori, H.~H. Patel, S.~Profumo, and D.~Tuckler, ``{Electric
  dipole moments in a leptoquark scenario for the $B$-physics anomalies},''
  \href{http://dx.doi.org/10.1007/JHEP05(2020)069}{{\em JHEP} {\bfseries 05}
  (2020) 069}, \href{http://arxiv.org/abs/2002.01400}{{\ttfamily
  arXiv:2002.01400 [hep-ph]}}.

\bibitem{Babu:2020hun}
K.~S. Babu, P.~S.~B. Dev, S.~Jana, and A.~Thapa, ``{Unified framework for
  $B$-anomalies, muon $g-2$ and neutrino masses},''
  \href{http://dx.doi.org/10.1007/JHEP03(2021)179}{{\em JHEP} {\bfseries 03}
  (2021) 179}, \href{http://arxiv.org/abs/2009.01771}{{\ttfamily
  arXiv:2009.01771 [hep-ph]}}.

\bibitem{Crivellin:2021rbq}
A.~Crivellin and M.~Hoferichter, ``{Consequences of chirally enhanced
  explanations of (g-2)$_{\mu}$ for h $\to$ \ensuremath{\mu}\ensuremath{\mu}
  and Z $\to$ \ensuremath{\mu}\ensuremath{\mu}},''
  \href{http://dx.doi.org/10.1007/JHEP07(2021)135}{{\em JHEP} {\bfseries 07}
  (2021) 135}, \href{http://arxiv.org/abs/2104.03202}{{\ttfamily
  arXiv:2104.03202 [hep-ph]}}. [Erratum: JHEP 10, 030 (2022)].

\bibitem{Crivellin:2018qmi}
A.~Crivellin, M.~Hoferichter, and P.~Schmidt-Wellenburg, ``{Combined
  explanations of $(g-2)_{\mu,e}$ and implications for a large muon EDM},''
  \href{http://dx.doi.org/10.1103/PhysRevD.98.113002}{{\em Phys. Rev. D}
  {\bfseries 98} no.~11, (2018) 113002},
  \href{http://arxiv.org/abs/1807.11484}{{\ttfamily arXiv:1807.11484
  [hep-ph]}}.

\bibitem{Hiller:2020fbu}
G.~Hiller, C.~Hormigos-Feliu, D.~F. Litim, and T.~Steudtner, ``{Model Building
  from Asymptotic Safety with Higgs and Flavor Portals},''
  \href{http://dx.doi.org/10.1103/PhysRevD.102.095023}{{\em Phys. Rev. D}
  {\bfseries 102} no.~9, (2020) 095023},
  \href{http://arxiv.org/abs/2008.08606}{{\ttfamily arXiv:2008.08606
  [hep-ph]}}.

\bibitem{Hamaguchi:2022byw}
K.~Hamaguchi, N.~Nagata, G.~Osaki, and S.-Y. Tseng, ``{Probing New Physics in
  the Vector-like Lepton Model by Lepton Electric Dipole Moments},''
  \href{http://arxiv.org/abs/2211.16800}{{\ttfamily arXiv:2211.16800
  [hep-ph]}}.

\bibitem{Yin:2021yqy}
W.~Yin, ``{Radiative lepton mass and muon g-2 with suppressed lepton flavor and
  CP violations},'' \href{http://dx.doi.org/10.1007/JHEP08(2021)043}{{\em JHEP}
  {\bfseries 08} (2021) 043}, \href{http://arxiv.org/abs/2103.14234}{{\ttfamily
  arXiv:2103.14234 [hep-ph]}}.

\bibitem{Baker:2021yli}
M.~J. Baker, P.~Cox, and R.~R. Volkas, ``{Radiative muon mass models and
  $(g-2)_\mu$},'' \href{http://dx.doi.org/10.1007/JHEP05(2021)174}{{\em JHEP}
  {\bfseries 05} (2021) 174}, \href{http://arxiv.org/abs/2103.13401}{{\ttfamily
  arXiv:2103.13401 [hep-ph]}}.

\bibitem{Buchmuller:2005eh}
W.~Buchmuller, R.~D. Peccei, and T.~Yanagida, ``{Leptogenesis as the origin of
  matter},''
  \href{http://dx.doi.org/10.1146/annurev.nucl.55.090704.151558}{{\em Ann. Rev.
  Nucl. Part. Sci.} {\bfseries 55} (2005) 311--355},
  \href{http://arxiv.org/abs/hep-ph/0502169}{{\ttfamily arXiv:hep-ph/0502169}}.

\bibitem{Hahn:1998yk}
T.~Hahn and M.~Perez-Victoria, ``{Automatized one loop calculations in
  four-dimensions and D-dimensions},''
  \href{http://dx.doi.org/10.1016/S0010-4655(98)00173-8}{{\em Comput. Phys.
  Commun.} {\bfseries 118} (1999) 153--165},
  \href{http://arxiv.org/abs/hep-ph/9807565}{{\ttfamily arXiv:hep-ph/9807565}}.

\bibitem{ATLAS:2020fzp}
{\bfseries ATLAS} Collaboration, G.~Aad {\em et~al.}, ``{A search for the
  dimuon decay of the Standard Model Higgs boson with the ATLAS detector},''
  \href{http://dx.doi.org/10.1016/j.physletb.2020.135980}{{\em Phys. Lett. B}
  {\bfseries 812} (2021) 135980},
  \href{http://arxiv.org/abs/2007.07830}{{\ttfamily arXiv:2007.07830
  [hep-ex]}}.

\bibitem{CMS:2020xwi}
{\bfseries CMS} Collaboration, A.~M. Sirunyan {\em et~al.}, ``{Evidence for
  Higgs boson decay to a pair of muons},''
  \href{http://dx.doi.org/10.1007/JHEP01(2021)148}{{\em JHEP} {\bfseries 01}
  (2021) 148}, \href{http://arxiv.org/abs/2009.04363}{{\ttfamily
  arXiv:2009.04363 [hep-ex]}}.

\bibitem{LHCHiggsCrossSectionWorkingGroup:2016ypw}
{\bfseries LHC Higgs Cross Section Working Group} Collaboration, D.~de~Florian
  {\em et~al.}, ``{Handbook of LHC Higgs Cross Sections: 4. Deciphering the
  Nature of the Higgs Sector},''
  \href{http://arxiv.org/abs/1610.07922}{{\ttfamily arXiv:1610.07922
  [hep-ph]}}.

\bibitem{ALEPH:2005ab}
{\bfseries ALEPH, DELPHI, L3, OPAL, SLD, LEP Electroweak Working Group, SLD
  Electroweak Group, SLD Heavy Flavour Group} Collaboration, S.~Schael {\em
  et~al.}, ``{Precision electroweak measurements on the $Z$ resonance},''
  \href{http://dx.doi.org/10.1016/j.physrep.2005.12.006}{{\em Phys. Rept.}
  {\bfseries 427} (2006) 257--454},
  \href{http://arxiv.org/abs/hep-ex/0509008}{{\ttfamily arXiv:hep-ex/0509008}}.

\bibitem{Baker:2020vkh}
M.~J. Baker, P.~Cox, and R.~R. Volkas, ``{Has the Origin of the Third-Family
  Fermion Masses been Determined?},''
  \href{http://dx.doi.org/10.1007/JHEP04(2021)151}{{\em JHEP} {\bfseries 04}
  (2021) 151}, \href{http://arxiv.org/abs/2012.10458}{{\ttfamily
  arXiv:2012.10458 [hep-ph]}}.

\bibitem{Griest:1990kh}
K.~Griest and D.~Seckel, ``{Three exceptions in the calculation of relic
  abundances},'' \href{http://dx.doi.org/10.1103/PhysRevD.43.3191}{{\em Phys.
  Rev. D} {\bfseries 43} (1991) 3191--3203}.

\bibitem{Gondolo:1990dk}
P.~Gondolo and G.~Gelmini, ``{Cosmic abundances of stable particles: Improved
  analysis},'' \href{http://dx.doi.org/10.1016/0550-3213(91)90438-4}{{\em Nucl.
  Phys. B} {\bfseries 360} (1991) 145--179}.

\bibitem{Edsjo:1997bg}
J.~Edsjo and P.~Gondolo, ``{Neutralino relic density including
  coannihilations},'' \href{http://dx.doi.org/10.1103/PhysRevD.56.1879}{{\em
  Phys. Rev. D} {\bfseries 56} (1997) 1879--1894},
  \href{http://arxiv.org/abs/hep-ph/9704361}{{\ttfamily arXiv:hep-ph/9704361}}.

\bibitem{Belanger:2018ccd}
G.~B\'elanger, F.~Boudjema, A.~Goudelis, A.~Pukhov, and B.~Zaldivar,
  ``{micrOMEGAs5.0 : Freeze-in},''
  \href{http://dx.doi.org/10.1016/j.cpc.2018.04.027}{{\em Comput. Phys.
  Commun.} {\bfseries 231} (2018) 173--186},
  \href{http://arxiv.org/abs/1801.03509}{{\ttfamily arXiv:1801.03509
  [hep-ph]}}.

\bibitem{Belanger:2020gnr}
G.~Belanger, A.~Mjallal, and A.~Pukhov, ``{Recasting direct detection limits
  within micrOMEGAs and implication for non-standard Dark Matter scenarios},''
  \href{http://dx.doi.org/10.1140/epjc/s10052-021-09012-z}{{\em Eur. Phys. J.
  C} {\bfseries 81} no.~3, (2021) 239},
  \href{http://arxiv.org/abs/2003.08621}{{\ttfamily arXiv:2003.08621
  [hep-ph]}}.

\bibitem{DelNobile:2021wmp}
E.~Del~Nobile, ``{The Theory of Direct Dark Matter Detection: A Guide to
  Computations},'' \href{http://arxiv.org/abs/2104.12785}{{\ttfamily
  arXiv:2104.12785 [hep-ph]}}.

\bibitem{Ellis:2000ds}
J.~R. Ellis, A.~Ferstl, and K.~A. Olive, ``{Reevaluation of the elastic
  scattering of supersymmetric dark matter},''
  \href{http://dx.doi.org/10.1016/S0370-2693(00)00459-7}{{\em Phys. Lett. B}
  {\bfseries 481} (2000) 304--314},
  \href{http://arxiv.org/abs/hep-ph/0001005}{{\ttfamily arXiv:hep-ph/0001005}}.

\bibitem{Gondolo:2004sc}
P.~Gondolo, J.~Edsjo, P.~Ullio, L.~Bergstrom, M.~Schelke, and E.~A. Baltz,
  ``{DarkSUSY: Computing supersymmetric dark matter properties numerically},''
  \href{http://dx.doi.org/10.1088/1475-7516/2004/07/008}{{\em JCAP} {\bfseries
  07} (2004) 008}, \href{http://arxiv.org/abs/astro-ph/0406204}{{\ttfamily
  arXiv:astro-ph/0406204}}.

\bibitem{Ellis:2008hf}
J.~R. Ellis, K.~A. Olive, and C.~Savage, ``{Hadronic Uncertainties in the
  Elastic Scattering of Supersymmetric Dark Matter},''
  \href{http://dx.doi.org/10.1103/PhysRevD.77.065026}{{\em Phys. Rev. D}
  {\bfseries 77} (2008) 065026},
  \href{http://arxiv.org/abs/0801.3656}{{\ttfamily arXiv:0801.3656 [hep-ph]}}.

\bibitem{Belanger:2008sj}
G.~Belanger, F.~Boudjema, A.~Pukhov, and A.~Semenov, ``{Dark matter direct
  detection rate in a generic model with micrOMEGAs 2.2},''
  \href{http://dx.doi.org/10.1016/j.cpc.2008.11.019}{{\em Comput. Phys.
  Commun.} {\bfseries 180} (2009) 747--767},
  \href{http://arxiv.org/abs/0803.2360}{{\ttfamily arXiv:0803.2360 [hep-ph]}}.

\bibitem{Cheng:2012qr}
H.-Y. Cheng and C.-W. Chiang, ``{Revisiting Scalar and Pseudoscalar Couplings
  with Nucleons},'' \href{http://dx.doi.org/10.1007/JHEP07(2012)009}{{\em JHEP}
  {\bfseries 07} (2012) 009}, \href{http://arxiv.org/abs/1202.1292}{{\ttfamily
  arXiv:1202.1292 [hep-ph]}}.

\bibitem{Belanger:2013oya}
G.~Belanger, F.~Boudjema, A.~Pukhov, and A.~Semenov, ``{micrOMEGAs$\_$3: A
  program for calculating dark matter observables},''
  \href{http://dx.doi.org/10.1016/j.cpc.2013.10.016}{{\em Comput. Phys.
  Commun.} {\bfseries 185} (2014) 960--985},
  \href{http://arxiv.org/abs/1305.0237}{{\ttfamily arXiv:1305.0237 [hep-ph]}}.

\bibitem{Crivellin:2013ipa}
A.~Crivellin, M.~Hoferichter, and M.~Procura, ``{Accurate evaluation of
  hadronic uncertainties in spin-independent WIMP-nucleon scattering:
  Disentangling two- and three-flavor effects},''
  \href{http://dx.doi.org/10.1103/PhysRevD.89.054021}{{\em Phys. Rev. D}
  {\bfseries 89} (2014) 054021},
  \href{http://arxiv.org/abs/1312.4951}{{\ttfamily arXiv:1312.4951 [hep-ph]}}.

\bibitem{Hoferichter:2015dsa}
M.~Hoferichter, J.~Ruiz~de Elvira, B.~Kubis, and U.-G. Mei\ss{}ner,
  ``{High-Precision Determination of the Pion-Nucleon \ensuremath{\sigma} Term
  from Roy-Steiner Equations},''
  \href{http://dx.doi.org/10.1103/PhysRevLett.115.092301}{{\em Phys. Rev.
  Lett.} {\bfseries 115} (2015) 092301},
  \href{http://arxiv.org/abs/1506.04142}{{\ttfamily arXiv:1506.04142
  [hep-ph]}}.

\bibitem{Ellis:2018dmb}
J.~Ellis, N.~Nagata, and K.~A. Olive, ``{Uncertainties in WIMP Dark Matter
  Scattering Revisited},''
  \href{http://dx.doi.org/10.1140/epjc/s10052-018-6047-y}{{\em Eur. Phys. J. C}
  {\bfseries 78} no.~7, (2018) 569},
  \href{http://arxiv.org/abs/1805.09795}{{\ttfamily arXiv:1805.09795
  [hep-ph]}}.

\bibitem{Helm:1956zz}
R.~H. Helm, ``{Inelastic and Elastic Scattering of 187-Mev Electrons from
  Selected Even-Even Nuclei},''
  \href{http://dx.doi.org/10.1103/PhysRev.104.1466}{{\em Phys. Rev.} {\bfseries
  104} (1956) 1466--1475}.

\bibitem{Lewin:1995rx}
J.~D. Lewin and P.~F. Smith, ``{Review of mathematics, numerical factors, and
  corrections for dark matter experiments based on elastic nuclear recoil},''
  \href{http://dx.doi.org/10.1016/S0927-6505(96)00047-3}{{\em Astropart. Phys.}
  {\bfseries 6} (1996) 87--112}.

\bibitem{LZ:2022ufs}
{\bfseries LZ} Collaboration, J.~Aalbers {\em et~al.}, ``{First Dark Matter
  Search Results from the LUX-ZEPLIN (LZ) Experiment},''
  \href{http://arxiv.org/abs/2207.03764}{{\ttfamily arXiv:2207.03764
  [hep-ex]}}.

\bibitem{XENON:2018voc}
{\bfseries XENON} Collaboration, E.~Aprile {\em et~al.}, ``{Dark Matter Search
  Results from a One Ton-Year Exposure of XENON1T},''
  \href{http://dx.doi.org/10.1103/PhysRevLett.121.111302}{{\em Phys. Rev.
  Lett.} {\bfseries 121} no.~11, (2018) 111302},
  \href{http://arxiv.org/abs/1805.12562}{{\ttfamily arXiv:1805.12562
  [astro-ph.CO]}}.

\bibitem{XENON:2019rxp}
{\bfseries XENON} Collaboration, E.~Aprile {\em et~al.}, ``{Constraining the
  spin-dependent WIMP-nucleon cross sections with XENON1T},''
  \href{http://dx.doi.org/10.1103/PhysRevLett.122.141301}{{\em Phys. Rev.
  Lett.} {\bfseries 122} no.~14, (2019) 141301},
  \href{http://arxiv.org/abs/1902.03234}{{\ttfamily arXiv:1902.03234
  [astro-ph.CO]}}.

\bibitem{PandaX-4T:2021bab}
{\bfseries PandaX-4T} Collaboration, Y.~Meng {\em et~al.}, ``{Dark Matter
  Search Results from the PandaX-4T Commissioning Run},''
  \href{http://dx.doi.org/10.1103/PhysRevLett.127.261802}{{\em Phys. Rev.
  Lett.} {\bfseries 127} no.~26, (2021) 261802},
  \href{http://arxiv.org/abs/2107.13438}{{\ttfamily arXiv:2107.13438
  [hep-ex]}}.

\bibitem{Liu:2022zgu}
{\bfseries PandaX} Collaboration, J.~Liu, ``{The first results of PandaX-4T},''
  \href{http://dx.doi.org/10.1142/S0218271822300075}{{\em Int. J. Mod. Phys. D}
  {\bfseries 31} no.~05, (2022) 2230007}.

\bibitem{Pospelov:2000bq}
M.~Pospelov and T.~ter Veldhuis, ``{Direct and indirect limits on the
  electromagnetic form-factors of WIMPs},''
  \href{http://dx.doi.org/10.1016/S0370-2693(00)00358-0}{{\em Phys. Lett. B}
  {\bfseries 480} (2000) 181--186},
  \href{http://arxiv.org/abs/hep-ph/0003010}{{\ttfamily arXiv:hep-ph/0003010}}.

\bibitem{Ho:2012bg}
C.~M. Ho and R.~J. Scherrer, ``{Anapole Dark Matter},''
  \href{http://dx.doi.org/10.1016/j.physletb.2013.04.039}{{\em Phys. Lett. B}
  {\bfseries 722} (2013) 341--346},
  \href{http://arxiv.org/abs/1211.0503}{{\ttfamily arXiv:1211.0503 [hep-ph]}}.

\bibitem{Lee:1977yc}
B.~W. Lee, C.~Quigg, and H.~B. Thacker, ``{The Strength of Weak Interactions at
  Very High-Energies and the Higgs Boson Mass},''
  \href{http://dx.doi.org/10.1103/PhysRevLett.38.883}{{\em Phys. Rev. Lett.}
  {\bfseries 38} (1977) 883--885}.

\bibitem{Lee:1977eg}
B.~W. Lee, C.~Quigg, and H.~B. Thacker, ``{Weak Interactions at Very
  High-Energies: The Role of the Higgs Boson Mass},''
  \href{http://dx.doi.org/10.1103/PhysRevD.16.1519}{{\em Phys. Rev. D}
  {\bfseries 16} (1977) 1519}.

\bibitem{Cynolter:2004cq}
G.~Cynolter, E.~Lendvai, and G.~Pocsik, ``{Note on unitarity constraints in a
  model for a singlet scalar dark matter candidate},'' {\em Acta Phys. Polon.
  B} {\bfseries 36} (2005) 827--832,
  \href{http://arxiv.org/abs/hep-ph/0410102}{{\ttfamily arXiv:hep-ph/0410102}}.

\bibitem{Kang:2013zba}
S.~K. Kang and J.~Park, ``{Unitarity Constraints in the standard model with a
  singlet scalar field},''
  \href{http://dx.doi.org/10.1007/JHEP04(2015)009}{{\em JHEP} {\bfseries 04}
  (2015) 009}, \href{http://arxiv.org/abs/1306.6713}{{\ttfamily arXiv:1306.6713
  [hep-ph]}}.

\bibitem{Costa:2014qga}
R.~Costa, A.~P. Morais, M.~O.~P. Sampaio, and R.~Santos, ``{Two-loop stability
  of a complex singlet extended Standard Model},''
  \href{http://dx.doi.org/10.1103/PhysRevD.92.025024}{{\em Phys. Rev. D}
  {\bfseries 92} (2015) 025024},
  \href{http://arxiv.org/abs/1411.4048}{{\ttfamily arXiv:1411.4048 [hep-ph]}}.

\bibitem{Casalbuoni:1986hy}
R.~Casalbuoni, D.~Dominici, R.~Gatto, and C.~Giunti, ``{Strong Interacting Two
  Doublet and Doublet Singlet Higgs Models},''
  \href{http://dx.doi.org/10.1016/0370-2693(86)91502-9}{{\em Phys. Lett. B}
  {\bfseries 178} (1986) 235}.

\bibitem{Casalbuoni:1987eg}
R.~Casalbuoni, D.~Dominici, F.~Feruglio, and R.~Gatto, ``{Testing the Standard
  Model in Terms of a Possible Strong Scalar Sector},''
  \href{http://dx.doi.org/10.1016/0370-2693(88)90158-X}{{\em Phys. Lett. B}
  {\bfseries 200} (1988) 495--500}.

\bibitem{Maalampi:1991fb}
J.~Maalampi, J.~Sirkka, and I.~Vilja, ``{Tree level unitarity and triviality
  bounds for two Higgs models},''
  \href{http://dx.doi.org/10.1016/0370-2693(91)90068-2}{{\em Phys. Lett. B}
  {\bfseries 265} (1991) 371--376}.

\bibitem{Kanemura:1993hm}
S.~Kanemura, T.~Kubota, and E.~Takasugi, ``{Lee-Quigg-Thacker bounds for Higgs
  boson masses in a two doublet model},''
  \href{http://dx.doi.org/10.1016/0370-2693(93)91205-2}{{\em Phys. Lett. B}
  {\bfseries 313} (1993) 155--160},
  \href{http://arxiv.org/abs/hep-ph/9303263}{{\ttfamily arXiv:hep-ph/9303263}}.

\bibitem{Ginzburg:2003fe}
I.~F. Ginzburg and I.~P. Ivanov, ``{Tree level unitarity constraints in the
  2HDM with CP violation},''
  \href{http://arxiv.org/abs/hep-ph/0312374}{{\ttfamily arXiv:hep-ph/0312374}}.

\bibitem{Akeroyd:2000wc}
A.~G. Akeroyd, A.~Arhrib, and E.-M. Naimi, ``{Note on tree level unitarity in
  the general two Higgs doublet model},''
  \href{http://dx.doi.org/10.1016/S0370-2693(00)00962-X}{{\em Phys. Lett. B}
  {\bfseries 490} (2000) 119--124},
  \href{http://arxiv.org/abs/hep-ph/0006035}{{\ttfamily arXiv:hep-ph/0006035}}.

\bibitem{Horejsi:2005da}
J.~Horejsi and M.~Kladiva, ``{Tree-unitarity bounds for THDM Higgs masses
  revisited},'' \href{http://dx.doi.org/10.1140/epjc/s2006-02472-3}{{\em Eur.
  Phys. J. C} {\bfseries 46} (2006) 81--91},
  \href{http://arxiv.org/abs/hep-ph/0510154}{{\ttfamily arXiv:hep-ph/0510154}}.

\bibitem{Porod:2003um}
W.~Porod, ``{SPheno, a program for calculating supersymmetric spectra, SUSY
  particle decays and SUSY particle production at e+ e- colliders},''
  \href{http://dx.doi.org/10.1016/S0010-4655(03)00222-4}{{\em Comput. Phys.
  Commun.} {\bfseries 153} (2003) 275--315},
  \href{http://arxiv.org/abs/hep-ph/0301101}{{\ttfamily arXiv:hep-ph/0301101}}.

\bibitem{Porod:2011nf}
W.~Porod and F.~Staub, ``{SPheno 3.1: Extensions including flavour, CP-phases
  and models beyond the MSSM},''
  \href{http://dx.doi.org/10.1016/j.cpc.2012.05.021}{{\em Comput. Phys.
  Commun.} {\bfseries 183} (2012) 2458--2469},
  \href{http://arxiv.org/abs/1104.1573}{{\ttfamily arXiv:1104.1573 [hep-ph]}}.

\bibitem{Staub:2008uz}
F.~Staub, ``{SARAH},'' \href{http://arxiv.org/abs/0806.0538}{{\ttfamily
  arXiv:0806.0538 [hep-ph]}}.

\bibitem{Staub:2009bi}
F.~Staub, ``{From Superpotential to Model Files for FeynArts and
  CalcHep/CompHep},'' \href{http://dx.doi.org/10.1016/j.cpc.2010.01.011}{{\em
  Comput. Phys. Commun.} {\bfseries 181} (2010) 1077--1086},
  \href{http://arxiv.org/abs/0909.2863}{{\ttfamily arXiv:0909.2863 [hep-ph]}}.

\bibitem{Staub:2010jh}
F.~Staub, ``{Automatic Calculation of supersymmetric Renormalization Group
  Equations and Self Energies},''
  \href{http://dx.doi.org/10.1016/j.cpc.2010.11.030}{{\em Comput. Phys.
  Commun.} {\bfseries 182} (2011) 808--833},
  \href{http://arxiv.org/abs/1002.0840}{{\ttfamily arXiv:1002.0840 [hep-ph]}}.

\bibitem{Staub:2012pb}
F.~Staub, ``{SARAH 3.2: Dirac Gauginos, UFO output, and more},''
  \href{http://dx.doi.org/10.1016/j.cpc.2013.02.019}{{\em Comput. Phys.
  Commun.} {\bfseries 184} (2013) 1792--1809},
  \href{http://arxiv.org/abs/1207.0906}{{\ttfamily arXiv:1207.0906 [hep-ph]}}.

\bibitem{Staub:2013tta}
F.~Staub, ``{SARAH 4 : A tool for (not only SUSY) model builders},''
  \href{http://dx.doi.org/10.1016/j.cpc.2014.02.018}{{\em Comput. Phys.
  Commun.} {\bfseries 185} (2014) 1773--1790},
  \href{http://arxiv.org/abs/1309.7223}{{\ttfamily arXiv:1309.7223 [hep-ph]}}.

\bibitem{Goodsell:2018tti}
M.~D. Goodsell and F.~Staub, ``{Unitarity constraints on general scalar
  couplings with SARAH},''
  \href{http://dx.doi.org/10.1140/epjc/s10052-018-6127-z}{{\em Eur. Phys. J. C}
  {\bfseries 78} no.~8, (2018) 649},
  \href{http://arxiv.org/abs/1805.07306}{{\ttfamily arXiv:1805.07306
  [hep-ph]}}.

\bibitem{Peskin:1990zt}
M.~E. Peskin and T.~Takeuchi, ``{A New constraint on a strongly interacting
  Higgs sector},'' \href{http://dx.doi.org/10.1103/PhysRevLett.65.964}{{\em
  Phys. Rev. Lett.} {\bfseries 65} (1990) 964--967}.

\bibitem{Peskin:1991sw}
M.~E. Peskin and T.~Takeuchi, ``{Estimation of oblique electroweak
  corrections},'' \href{http://dx.doi.org/10.1103/PhysRevD.46.381}{{\em Phys.
  Rev. D} {\bfseries 46} (1992) 381--409}.

\bibitem{Grimus:2007if}
W.~Grimus, L.~Lavoura, O.~M. Ogreid, and P.~Osland, ``{A Precision constraint
  on multi-Higgs-doublet models},''
  \href{http://dx.doi.org/10.1088/0954-3899/35/7/075001}{{\em J. Phys. G}
  {\bfseries 35} (2008) 075001},
  \href{http://arxiv.org/abs/0711.4022}{{\ttfamily arXiv:0711.4022 [hep-ph]}}.

\bibitem{Grimus:2008nb}
W.~Grimus, L.~Lavoura, O.~M. Ogreid, and P.~Osland, ``{The Oblique parameters
  in multi-Higgs-doublet models},''
  \href{http://dx.doi.org/10.1016/j.nuclphysb.2008.04.019}{{\em Nucl. Phys. B}
  {\bfseries 801} (2008) 81--96},
  \href{http://arxiv.org/abs/0802.4353}{{\ttfamily arXiv:0802.4353 [hep-ph]}}.

\bibitem{ParticleDataGroup:2022pth}
{\bfseries Particle Data Group} Collaboration, R.~L. Workman {\em et~al.},
  ``{Review of Particle Physics},''
  \href{http://dx.doi.org/10.1093/ptep/ptac097}{{\em PTEP} {\bfseries 2022}
  (2022) 083C01}.

\bibitem{Planck:2018vyg}
{\bfseries Planck} Collaboration, N.~Aghanim {\em et~al.}, ``{Planck 2018
  results. VI. Cosmological parameters},''
  \href{http://dx.doi.org/10.1051/0004-6361/201833910}{{\em Astron. Astrophys.}
  {\bfseries 641} (2020) A6}, \href{http://arxiv.org/abs/1807.06209}{{\ttfamily
  arXiv:1807.06209 [astro-ph.CO]}}. [Erratum: Astron.Astrophys. 652, C4
  (2021)].

\bibitem{Pontecorvo:1957cp}
B.~Pontecorvo, ``{Mesonium and anti-mesonium},'' {\em Sov. Phys. JETP}
  {\bfseries 6} (1957) 429.

\bibitem{Feinberg:1961zza}
G.~Feinberg and S.~Weinberg, ``{Conversion of Muonium into Antimuonium},''
  \href{http://dx.doi.org/10.1103/PhysRev.123.1439}{{\em Phys. Rev.} {\bfseries
  123} (1961) 1439--1443}.

\bibitem{Lee:1977tib}
B.~W. Lee and R.~E. Shrock, ``{Natural Suppression of Symmetry Violation in
  Gauge Theories: Muon - Lepton and Electron Lepton Number Nonconservation},''
  \href{http://dx.doi.org/10.1103/PhysRevD.16.1444}{{\em Phys. Rev. D}
  {\bfseries 16} (1977) 1444}.

\bibitem{Lee:1977qz}
B.~W. Lee, S.~Pakvasa, R.~E. Shrock, and H.~Sugawara, ``{Muon and Electron
  Number Nonconservation in a V-A Gauge Model},''
  \href{http://dx.doi.org/10.1103/PhysRevLett.38.937}{{\em Phys. Rev. Lett.}
  {\bfseries 38} (1977) 937}. [Erratum: Phys.Rev.Lett. 38, 1230 (1977)].

\bibitem{Willmann:1998gd}
L.~Willmann {\em et~al.}, ``{New bounds from searching for muonium to
  anti-muonium conversion},''
  \href{http://dx.doi.org/10.1103/PhysRevLett.82.49}{{\em Phys. Rev. Lett.}
  {\bfseries 82} (1999) 49--52},
  \href{http://arxiv.org/abs/hep-ex/9807011}{{\ttfamily arXiv:hep-ex/9807011}}.

\bibitem{deBlas:2019rxi}
J.~de~Blas {\em et~al.}, ``{Higgs Boson Studies at Future Particle
  Colliders},'' \href{http://dx.doi.org/10.1007/JHEP01(2020)139}{{\em JHEP}
  {\bfseries 01} (2020) 139}, \href{http://arxiv.org/abs/1905.03764}{{\ttfamily
  arXiv:1905.03764 [hep-ph]}}.

\bibitem{ATLAS:2019lff}
{\bfseries ATLAS} Collaboration, G.~Aad {\em et~al.}, ``{Search for electroweak
  production of charginos and sleptons decaying into final states with two
  leptons and missing transverse momentum in $\sqrt{s}=13$ TeV $pp$ collisions
  using the ATLAS detector},''
  \href{http://dx.doi.org/10.1140/epjc/s10052-019-7594-6}{{\em Eur. Phys. J. C}
  {\bfseries 80} no.~2, (2020) 123},
  \href{http://arxiv.org/abs/1908.08215}{{\ttfamily arXiv:1908.08215
  [hep-ex]}}.

\bibitem{ATLAS:2019lng}
{\bfseries ATLAS} Collaboration, G.~Aad {\em et~al.}, ``{Searches for
  electroweak production of supersymmetric particles with compressed mass
  spectra in $\sqrt{s}=$ 13 TeV $pp$ collisions with the ATLAS detector},''
  \href{http://dx.doi.org/10.1103/PhysRevD.101.052005}{{\em Phys. Rev. D}
  {\bfseries 101} no.~5, (2020) 052005},
  \href{http://arxiv.org/abs/1911.12606}{{\ttfamily arXiv:1911.12606
  [hep-ex]}}.

\bibitem{CMS:2018eqb}
{\bfseries CMS} Collaboration, A.~M. Sirunyan {\em et~al.}, ``{Search for
  supersymmetric partners of electrons and muons in proton-proton collisions at
  $\sqrt{s}=$ 13 TeV},''
  \href{http://dx.doi.org/10.1016/j.physletb.2019.01.005}{{\em Phys. Lett. B}
  {\bfseries 790} (2019) 140--166},
  \href{http://arxiv.org/abs/1806.05264}{{\ttfamily arXiv:1806.05264
  [hep-ex]}}.

\bibitem{Kawamura:2020qxo}
J.~Kawamura, S.~Okawa, and Y.~Omura, ``{Current status and muon $g-2$
  explanation of lepton portal dark matter},''
  \href{http://dx.doi.org/10.1007/JHEP08(2020)042}{{\em JHEP} {\bfseries 08}
  (2020) 042}, \href{http://arxiv.org/abs/2002.12534}{{\ttfamily
  arXiv:2002.12534 [hep-ph]}}.

\bibitem{Garny:2013ama}
M.~Garny, A.~Ibarra, M.~Pato, and S.~Vogl, ``{Internal bremsstrahlung
  signatures in light of direct dark matter searches},''
  \href{http://dx.doi.org/10.1088/1475-7516/2013/12/046}{{\em JCAP} {\bfseries
  12} (2013) 046}, \href{http://arxiv.org/abs/1306.6342}{{\ttfamily
  arXiv:1306.6342 [hep-ph]}}.

\bibitem{PAMELA:2010kea}
{\bfseries PAMELA} Collaboration, O.~Adriani {\em et~al.}, ``{PAMELA results on
  the cosmic-ray antiproton flux from 60 MeV to 180 GeV in kinetic energy},''
  \href{http://dx.doi.org/10.1103/PhysRevLett.105.121101}{{\em Phys. Rev.
  Lett.} {\bfseries 105} (2010) 121101},
  \href{http://arxiv.org/abs/1007.0821}{{\ttfamily arXiv:1007.0821
  [astro-ph.HE]}}.

\bibitem{Garny:2011cj}
M.~Garny, A.~Ibarra, and S.~Vogl, ``{Antiproton constraints on dark matter
  annihilations from internal electroweak bremsstrahlung},''
  \href{http://dx.doi.org/10.1088/1475-7516/2011/07/028}{{\em JCAP} {\bfseries
  07} (2011) 028}, \href{http://arxiv.org/abs/1105.5367}{{\ttfamily
  arXiv:1105.5367 [hep-ph]}}.

\bibitem{Garny:2011ii}
M.~Garny, A.~Ibarra, and S.~Vogl, ``{Dark matter annihilations into two light
  fermions and one gauge boson: General analysis and antiproton constraints},''
  \href{http://dx.doi.org/10.1088/1475-7516/2012/04/033}{{\em JCAP} {\bfseries
  04} (2012) 033}, \href{http://arxiv.org/abs/1112.5155}{{\ttfamily
  arXiv:1112.5155 [hep-ph]}}.

\bibitem{Mount:2017qzi}
B.~J. Mount {\em et~al.}, ``{LUX-ZEPLIN (LZ) Technical Design Report},''
  \href{http://arxiv.org/abs/1703.09144}{{\ttfamily arXiv:1703.09144
  [physics.ins-det]}}.

\bibitem{LZ:2019sgr}
{\bfseries LZ} Collaboration, D.~S. Akerib {\em et~al.}, ``{The LUX-ZEPLIN (LZ)
  Experiment},'' \href{http://dx.doi.org/10.1016/j.nima.2019.163047}{{\em Nucl.
  Instrum. Meth. A} {\bfseries 953} (2020) 163047},
  \href{http://arxiv.org/abs/1910.09124}{{\ttfamily arXiv:1910.09124
  [physics.ins-det]}}.

\bibitem{PandaX:2018wtu}
{\bfseries PandaX} Collaboration, H.~Zhang {\em et~al.}, ``{Dark matter direct
  search sensitivity of the PandaX-4T experiment},''
  \href{http://dx.doi.org/10.1007/s11433-018-9259-0}{{\em Sci. China Phys.
  Mech. Astron.} {\bfseries 62} no.~3, (2019) 31011},
  \href{http://arxiv.org/abs/1806.02229}{{\ttfamily arXiv:1806.02229
  [physics.ins-det]}}.

\bibitem{AlAli:2021let}
H.~Al~Ali {\em et~al.}, ``{The muon Smasher\textquoteright{}s guide},''
  \href{http://dx.doi.org/10.1088/1361-6633/ac6678}{{\em Rept. Prog. Phys.}
  {\bfseries 85} no.~8, (2022) 084201},
  \href{http://arxiv.org/abs/2103.14043}{{\ttfamily arXiv:2103.14043
  [hep-ph]}}.

\bibitem{Abe:2017jqo}
T.~Abe, R.~Sato, and K.~Yagyu, ``{Muon specific two-Higgs-doublet model},''
  \href{http://dx.doi.org/10.1007/JHEP07(2017)012}{{\em JHEP} {\bfseries 07}
  (2017) 012}, \href{http://arxiv.org/abs/1705.01469}{{\ttfamily
  arXiv:1705.01469 [hep-ph]}}.

\bibitem{Lindner:2016bgg}
M.~Lindner, M.~Platscher, and F.~S. Queiroz, ``{A Call for New Physics : The
  Muon Anomalous Magnetic Moment and Lepton Flavor Violation},''
  \href{http://dx.doi.org/10.1016/j.physrep.2017.12.001}{{\em Phys. Rept.}
  {\bfseries 731} (2018) 1--82},
  \href{http://arxiv.org/abs/1610.06587}{{\ttfamily arXiv:1610.06587
  [hep-ph]}}.

\bibitem{MEG:2016leq}
{\bfseries MEG} Collaboration, A.~M. Baldini {\em et~al.}, ``{Search for the
  lepton flavour violating decay $\mu ^+ \rightarrow \mathrm {e}^+ \gamma $
  with the full dataset of the MEG experiment},''
  \href{http://dx.doi.org/10.1140/epjc/s10052-016-4271-x}{{\em Eur. Phys. J. C}
  {\bfseries 76} no.~8, (2016) 434},
  \href{http://arxiv.org/abs/1605.05081}{{\ttfamily arXiv:1605.05081
  [hep-ex]}}.

\bibitem{Mertig:1990an}
R.~Mertig, M.~Bohm, and A.~Denner, ``{FEYN CALC: Computer algebraic calculation
  of Feynman amplitudes},''
  \href{http://dx.doi.org/10.1016/0010-4655(91)90130-D}{{\em Comput. Phys.
  Commun.} {\bfseries 64} (1991) 345--359}.

\bibitem{Shtabovenko:2016sxi}
V.~Shtabovenko, R.~Mertig, and F.~Orellana, ``{New Developments in FeynCalc
  9.0},'' \href{http://dx.doi.org/10.1016/j.cpc.2016.06.008}{{\em Comput. Phys.
  Commun.} {\bfseries 207} (2016) 432--444},
  \href{http://arxiv.org/abs/1601.01167}{{\ttfamily arXiv:1601.01167
  [hep-ph]}}.

\bibitem{Shtabovenko:2020gxv}
V.~Shtabovenko, R.~Mertig, and F.~Orellana, ``{FeynCalc 9.3: New features and
  improvements},'' \href{http://dx.doi.org/10.1016/j.cpc.2020.107478}{{\em
  Comput. Phys. Commun.} {\bfseries 256} (2020) 107478},
  \href{http://arxiv.org/abs/2001.04407}{{\ttfamily arXiv:2001.04407
  [hep-ph]}}.

\bibitem{tHooft:1972tcz}
G.~'t~Hooft and M.~J.~G. Veltman, ``{Regularization and Renormalization of
  Gauge Fields},'' \href{http://dx.doi.org/10.1016/0550-3213(72)90279-9}{{\em
  Nucl. Phys. B} {\bfseries 44} (1972) 189--213}.

\bibitem{Breitenlohner:1977hr}
P.~Breitenlohner and D.~Maison, ``{Dimensional Renormalization and the Action
  Principle},'' \href{http://dx.doi.org/10.1007/BF01609069}{{\em Commun. Math.
  Phys.} {\bfseries 52} (1977) 11--38}.

\end{thebibliography}\endgroup

\end{document}